\documentclass[preprint2]{proto} 
\usepackage{times}
\usepackage{color}
\usepackage{amsmath}
\usepackage{color}

\usepackage{natbib}
\bibpunct{(}{)}{;}{a}{,}{,}

\usepackage[]{todonotes}
\usepackage[squaren,textstyle]{SIunits}

\newcommand{\Cu}{\joule\usk\power{\kilogram}{-2}\usk\power{\kelvin}{-1}}
\newcommand{\Du}{\kilogram\usk\power{\metre}{-3}}
\newcommand{\tiu}{\joule\usk\power{\metre}{-2}\power{\second}{-1/2}\power{\kelvin}{-1}}
\newcommand{\Ku}{\watt\usk\power{\kelvin}{-1}\usk\power{\metre}{-1}}

\begin{document}

\title{\textbf{\LARGE Asteroid thermophysical modeling}}

\author {\textbf{\large Marco Delbo}}
%\affil{\small\em Lagrange Laboratory, University of Nice-Sophia, CNRS, C\^ote d'Azur Observatory}
\affil{\small\em Laboratoire Lagrange, UNS-CNRS, Observatoire de la C\^ote d'Azur}
\author {\textbf{\large Michael Mueller}}
\affil{\small\em SRON Netherlands Institute for Space Research}
\author {\textbf{\large Joshua P. Emery}}
\affil{\small\em Dept. of Earth and Planetary Sciences - University of Tennessee}
\author {\textbf{\large Ben Rozitis}}
\affil{\small\em Dept. of Earth and Planetary Sciences - University of Tennessee}
\author {\textbf{\large Maria Teresa Capria}}
\affil{\small\em Istituto di Astrofisica e Planetologia Spaziali, INAF}

\begin{abstract}
\begin{list}{ } {\rightmargin 1in}
%{\leftmargin -1in}
\baselineskip = 11pt
%rule{4.75in}{0.5pt}
%\vskip 1pt
\parindent=1pc
{\small 
The field of asteroid thermophysical modeling has experienced an extraordinary growth in the last ten years, %i.e. since \emph{Asteroids III}.  
as new thermal infrared data became available for hundreds of thousands of asteroids.  
The infrared emission of asteroids depends on the body's size, shape, albedo, thermal inertia, roughness and rotational properties. 
These parameters can therefore be derived by thermophysical modeling of infrared data. 
Thermophysical modeling led to asteroid size estimates that were confirmed at the few-percent level by later spacecraft visits. We discuss how instrumentation advances now allow mid-infrared interferometric observations as well as high-accuracy spectro-photometry, posing their own set of thermal-modeling challenges.
We present major breakthroughs achieved in studies of the thermal inertia, a sensitive indicator for the nature of asteroids soils, allowing us, for instance, to determine the grain size of asteroidal regoliths. Thermal inertia also governs non-gravitational  effects on asteroid orbits, requiring thermophysical modeling for precise asteroid dynamical studies. The radiative heating of asteroids, meteoroids, and comets from the Sun also governs the thermal stress in surface material; only recently has it been recognized as a significant weathering process.
%We now know the typical thermal inertia of small asteroids including NEAs. Their unexpectedly low thermal inertia implies that even small asteroids are typically covered in regolith, severely challenging previous expectations.
Asteroid space missions with thermal infrared instruments are currently undergoing study at all major space agencies. This will require a 
high level of sophistication of thermophysical models in order to analyze high-quality spacecraft data.
 \\~\\~\\~}%leave this in to get the correct vertical space after the abstract
\end{list}
\end{abstract}  

% !TEX root = astIVtpm_Rev2.tex
\section{\textbf{INTRODUCTION}}
 
Asteroid thermophysical modeling is about calculating the temperature of asteroids' surface and immediate subsurface, % by means of computer models 
%-- the so-called asteroid thermophyscal models (TPM) -- 
%that take into account a number of 
which depend on
absorption of sunlight, multiple scattering of reflected and thermally emitted photons, and  heat conduction. Physical parameters such as  albedo (or reflectivity), thermal conductivity, heat capacity, emissivity, density and roughness, along with the shape (e.g., elevation model) of the body, its orientation in space, and its previous thermal history are taken into account. 
% MIGO
From the synthetic surface temperatures, thermally emitted fluxes (typically in the infrared) can be calculated.  Physical properties are constrained by fitting model fluxes to observational data.
%MARCO
%Given surface temperatures, infrared fluxes can be calculated as a function of said physical parameters. Model fluxes are typically compared with measured ones in order to derive the values of physical characteristics of asteroids that give the best fit to the observational data. 

One differentiates between sophisticated \emph{thermophysical models} \citep[\emph{TPMs;}][]{Lebofsky1989aste.conf..128L,Spencer1990Icar...83...27S,Spencer1989Icar...78..337S,Lagerros1997A&A...325.1226L,Lagerros1996A&A...310.1011L,Lagerros1998A&A...332.1123L,Delbo2004PhD,Mueller2007PhD,Rozitis2011MNRAS.415.2042R} and \emph{simple thermal models}, which typically assume spherical shape, neglect heat conduction (or simplify its treatment), and do not treat surface roughness \citep[see][for reviews]{Harris2002aste.conf..205H,Delbo2002M&PS...37.1929D}. 
%Simple thermal models are typically used to calculate temperatures for those asteroids whose shape and spin pole are not known (which are the large majority). 
%We do not discuss the simple thermal models in details this chapter: we just provide an introduction to the NEATM \citep{Harris1998Icar..131..291H}, which is today the default model to derive asteroid sizes. 
In the past, usage of TPMs was reserved to the few exceptional asteroids 
%While in the past TPMs could only be used for a handful of asteroids, 
for which detailed shape models and high quality thermal infrared data existed \citep{Harris2002aste.conf..205H}.  In the last ten years, however, TPMs became significantly more applicable (see \S~\ref{S:results}), thanks both to
new spaceborn infrared telescopes
\citep[\emph{Spitzer}, \emph{WISE} and \emph{AKARI;} see][]{Mainzer2014astIV} and to the availability of an ever-growing number of asteroid shape models  \citep{Durech2014astIV}.%: For instance, at the time of writing, the DAMIT database (\url{http://astro.troja.mff.cuni.cz/projects/asteroids3D/web.ph}) contains 650 models for 382 asteroids, most of which have thermal IR data \citep[see also][]{Durech2014astIV}.

After introducing the motivations and the different contexts for calculating asteroid temperatures (\S~\ref{S:motivations}), we provide an overview of simple thermal models (\S~\ref{S:simpleTM}) and of TPMs (\S~\ref{S:tpm}).  We describe data analysis techniques based on TPMs (\S~\ref{S:implementationTPM}), then we present the latest results and implications on the physics of asteroids  (\S~\ref{S:results}).
In \S~\ref{S:tempEffect}, we discuss temperature-induced surface changes on asteroids;
see also the chapter "Asteroid surface geophysics" by \citet{Murdoch2014astIV}. All used symbols are summarized in Tab.~\ref{T:Nomenclature}. 

Note that we do not discuss here the so-called asteroid \emph{thermal evolution models}  that are generally used to compute the temperature throughout the body as a function of time, typically taking into account internal heat sources such as the decay of the radiogenic $^{26}$Al. Such models allow one to estimate the degree of metamorphism, aqueous alteration, melting and differentiation that asteroids experienced during the early phases of the solar system formation \citep[see][for a review]{McSween2002aste.conf..559M}.

% !TEX root = astIVtpm_Rev2.tex
\section{\textbf{MOTIVATIONS AND APPLICATIONS OF TPMs}}
\label{S:motivations}

%% Migo, 2015/02/12: nothing's wrong with the following paragraph, it's just that it feels wrong to start this section with a negative statement.
%Simple thermal models have limitations when it comes to detailed physical interpretation of high-quality observational data or the prediction of accurate thermal infrared fluxes from asteroids for calibration and other purposes \citep{2002A&A...381..324M}.

Thermophysical modeling of observations of asteroids in the thermal infrared ($\lambda \gtrsim$ 4 \micron) is a powerful technique to determine the values of  physical parameters of asteroids such as their sizes \citep[e.g.,][]{Muller2014PASJ...66...52M}, the thermal inertia and the roughness of their soils \citep[e.g.,][]{Muller1998A&A...338..340M,Mueller2007PhD,Delbo2009P&SS...57..259D,Matter2011Icar..215...47M,Rozitis2014A&A...568A..43R,Capria2014GeoRL..41.1438C}  and in some particular cases also of their bulk density and their bulk porosity \citep{Rozitis2013A&A...555A..20R,Rozitis2014Natur.512..174R,Emery2014Icar..234...17E,Chesley2014Icar..235....5C}

Knowledge of physical properties is crucial to understand asteroids: for instance, size information is fundamental to constrain the asteroid size frequency distribution that informs us about the collisional evolution of these bodies \citep{Bottke2005Icar..175..111B}; is paramount for the study of asteroid families, for the Earth-impact risk assessment of near-Earth asteroids \citep[NEAs; see][for a review]{Harris2014astIV}, and for the development of asteroid space mission scenarios (\S~\ref{S:missions}). Accurate sizes are also a prerequisite to calculate the volumes of asteroids for which we know the mass, allowing us to derive the bulk density, which inform us about the internal structure of these bodies \citep[e.g.,][]{Carry2012P&SS...73...98C}. 

Thermal inertia, the resistance of a material to temperature change (\S~\ref{S:thermalInertia}), is a sensitive indicator for the properties of the grainy soil \citep[regolith,][]{Murdoch2014astIV} on asteroids, e.g.,
the typical grain size \citep{Gundlach2013Icar..223..479G} and their degree of cementation \citep{Piqueux2009JGRE..114.9005P,Piqueux2009JGRE..114.9006P} can be inferred from thermal-inertia measurements.
%
%on exposure of solid rocks and boulders within the top few centimeters of the subsurface of the body and the typical particle size \citep{Gundlach2013Icar..223..479G}, degree of compaction \citep{Piqueux2009JGRE..114.9006P,Piqueux2009JGRE..114.9005P}, and depth of the regolith. The latter is the granular material the constitute the soil of asteroids \citep{Murdoch2014astIV} resulting from the break up of rocks exposed on the asteroid surface to impacts of meteoroids  \citep{Horz1997M&PS...32..179H} and thermal cracking \citep{Delbo2014Natur.508..233D}.
%
In general, the regolith is what we study by means of remote-sensing observations. Understanding the regolith is therefore crucial to infer the nature of the underlying body. Regolith informs us about the geological processes occurring on asteroids \citep{Murdoch2014astIV} such as impacts, micrometeoroid bombardment \citep{Horz1997M&PS...32..179H}, and thermal cracking \citep{Delbo2014Natur.508..233D}. Regolith contains records of elements implanted by the solar wind and cosmic radiation, and therefore informs us about the sources of those materials \citep{Lucey2006JGRE..111.8003L}. Regolith porosity can shed light on the role of electrostatic and van-der-Waals forces acting on the surface of these bodies \citep[e.g.,][]{Rozitis2014Natur.512..174R, Vernazza2012Icar..221.1162V}.

Knowledge of surface temperatures is also essential for the design of the instruments and for the  near-surface operation of space missions, as in the case of the sample-return missions Hayabusa-II and OSIRIS-REx of JAXA and NASA, respectively, %\citep{Lauretta2012LPICo1667.6291L}. 
%For instance, in the case of NASA sample-return mission OSIRIS-REx \citep{Lauretta2012LPICo1667.6291L}, the requirement of sampling a regolith not hotter than $\sim$350 K poses constraints on the latitude of the sample selection area on the body, the time of (asteroidal) day, as well as on the date of arrival at the asteroid. 
In the future, knowledge of asteroid temperatures will be crucial for planning human interaction with asteroids.

Another reason to model asteroid surface temperatures is that they affect its orbital and spin state evolution via the Yarkovsky and YORP effects, respectively \citep[\S~\ref{S:YarkoYORP} and ][]{Vokrouhlicky2014astIV}. 
In particular, thermal inertia  dictates the strength of the asteroid Yarkovsky effect. %, because it introduces a time-lag between absorption and re-radiation of solar radiation. 
This influences the dispersion of members of asteroid families, the orbital evolution of potentially hazardous asteroids, and the delivery of $D\lesssim$ 40 km asteroids and meteoroids from the main belt into dynamical resonance zones capable of transporting them to Earth-crossing orbits \citep[see][and the references therein]{Vokrouhlicky2014astIV}.

The YORP effect is believed to be shaping the distribution of rotation rates \citep{Bottke2006AREPS..34..157B} and spin vector orientation \citep{Vokrouhlicky2003Natur.425..147V,Hanus2011A&A...530A.134H,Hanus2013A&A...559A.134H}; small gravitationally bound aggregates could be spun up so fast \citep[and references therein]{Vokrouhlicky2014astIV,Bottke2006AREPS..34..157B} that they are forced to change shape and/or undergo mass shedding \citep{Holsapple2010Icar..205..430H}. Approximately 15\% of near-Earth asteroids are observed to be binaries \citep{Pravec2006Icar..181...63P}, and YORP spin up is proposed as a viable formation mechanism \citep{Walsh2008Natur.454..188W,Scheeres2007Icar..189..370S,Jacobson2011Icar..214..161J}. 

A further motivation to apply TPM techniques is to constrain the spin-axis orientation and the sense of rotation of asteroids 
\cite[examples are 101955 Bennu and 2005 YU$_\text{55}$,][]{Muller2012A&A...548A..36M,Muller2013A&A...558A..97M}. \cite{Durech2014astIV} describe  how to use optical and thermal infrared data simultaneously to derive more reliable asteroid shapes and spin properties.

%An important consequence of said thermal effects is that the density of asteroids can be derived by modeling accurate measurements of the orbital drift rate due to the Yarkovsky effect \citep{Emery2014Icar..234...17E,Rozitis2013A&A...555A..20R,Rozitis2014Natur.512..174R} and/or the acceleration of the rate of rotation due to the YORP effect \citep{Lowry2014A&A...562A..48L}

The temperature and its evolution through the entire life of an asteroid can alter its surface composition and nature of the regolith (\S~\ref{S:tempEffect}). For example, when the temperature rises above a certain threshold for a sustained period, certain volatiles can be lost  via  sublimation \citep{Schorghofer2008ApJ...682..697S,Capria2012A&A...537A..71C}, dehydration \citep{Marchi2009MNRAS.400..147M}, or desiccation \cite[][and references therein]{Delbo2011ApJ...728L..42D,Jewitt2014astIV}. 
%The threshold temperature for these processes depends on the presence of thermally insulating fine regolith protecting the volatile compounds underneath \citep[see e.g.,][for the case of water ice buried under a regolith layer]{Schorghofer2008ApJ...682..697S,Capria2012A&A...537A..71C}
%Note that processes such as dehydration or desiccation can be very important for NEAs due to their orbital evolution; temperature histories are therefore not trivially correlated with their present orbit \citep{Marchi2009MNRAS.400..147M}. 
%Certain near-Earth asteroids, such as (3200) Phaethon, have orbits that bring them close enough to the Sun that their surface temperature can exceed 1000 K! It is postulated that thermal desiccation and cracking, occurring at these temperatures, can drive the activity recently detected from this asteroid \citep{Jewitt2010AJ....140.1519J,Jewitt2014astIV}, but it is unclear wether it can explain the massive meteor stream of the Geminids, which are dynamically associated with Phaethon.

There can be pronounced and fast temperature variations between day and night. 
Modeling these temperature variations is fundamental to studying the effect of thermal cracking of asteroid surface material (\S~\ref{S:thermalCracking}), which was found to be an important source of fresh regolith production \citep{Delbo2014Natur.508..233D}.

\section{SIMPLE THERMAL MODELS}
\label{S:simpleTM}
We start by introducing the near-Earth asteroid thermal model \citep[NEATM,][]{Harris1998Icar..131..291H} that is typically used where the data quality and/or the available knowledge about asteroid shape and spin preclude the usage of TPMs.  Typically, the NEATM allows a robust estimation of asteroid diameter and albedo, but does not provide any direct information on thermal inertia or surface roughness \citep[see][for a review]{Harris2002aste.conf..205H}.
The recent large-scale thermal-emission surveys of asteroids and trans-Neptunian objects \citep[see][and references therein]{Mainzer2014astIV,Lellouch2013A&A...557A..60L} typically use the NEATM in their data analysis, thereby establishing it as the 
de-facto default among the simple thermal models. The typical NEATM accuracy is 15\% in diameter and roughly 30\% in albedo \citep{Harris2006IAUS..229..449H}. Other simple thermal models are the ``Standard Thermal Model'' \citep[STM;][]{Lebofsky1986Icar...68..239L}, the ``Isothermal Latitude Model'' \citep[ILM, also known as the Fast Rotating Model or FRM,][]{Lebofsky1989aste.conf..128L}, and the night emission simulated thermal model \cite[NESTM by][]{Wolters2009MNRAS.400..204W}. The STM and the ILM, reviewed by, e.g., \citet{Harris2002aste.conf..205H}, have largely fallen out of use.

The NEATM assumes that the asteroid has a spherical shape and does not directly account for thermal inertia nor surface roughness.  The surface temperature is given by the instantaneous equilibrium with the insolation, which is proportional to the cosine of the angular distance between local zenith and the Sun and zero at the night side. The maximum temperature occurs at the subsolar point and it reads:
\begin{equation}
(1-A)S_\odot r^{-2} = \eta \sigma \epsilon~T_\text{SS}^4
\label{E:equilib}
\end{equation}
(nomenclature is provided in Tab.~\ref{T:Nomenclature}).
The parameter $\eta$ was introduced in the STM of \citet{Lebofsky1986Icar...68..239L} as a means of changing the model temperature distribution to take account of the observed enhancement of thermal emission at small solar phase angles due to surface roughness. This is known as the beaming effect. For this reason $\eta$ is also called the \emph{beaming parameter}. 
The $\eta$ formalism, in the NEATM, allows a first-order description of the effect of a number of geometrical and physical parameters, in particular the thermal inertia and surface roughness on the spectral energy distribution of an asteroid \citep{Delbo2007Icar..190..236D}.
For a large thermal inertia, one would expect $\eta$-values significantly larger than unity \citep[e.g., 1.5--3; with theoretical maximum values around 3.5;][]{Delbo2007Icar..190..236D}, whereas for low thermal inertia $\eta \simeq$ 1 (for $\Gamma$ = 0  and zero surface roughness). Roughness, on the other hand, tends to lower the value of $\eta$ (for observations at low or moderate phase angles). For instance, a value of $\eta \sim$ 0.8 for a main belt asteroid indicates that this body has low thermal inertia and significant roughness \citep[with minimum theoretical values of 0.6 - 0.7;][]{Spencer1990Icar...83...27S,Delbo2007Icar..190..236D}. We note, however, that $\eta$ is not a physical property of an asteroid, as it can vary due to changing observing and illumination geometry, aspect angle, heliocentric distance of the body, phase angle and wavelength of observation.

% !TEX root = astIVtpm_Rev2.tex
\begin{deluxetable}{c l l || c l l }
\tabletypesize{\scriptsize}
\tablecaption{\label{T:Nomenclature}Nomenclature}
\tablewidth{0pt}
\tablehead{Symbol & Quantity & Unit & Symbol & Quantity & Unit}
\startdata
$T$                 & Temperature & K &                                                      $\bar{\theta}$      & Mean surface slope & deg \\                                     
$T_\text{SS}$       & Subsolar temperature & K &                                             $l_s$               & Thermal skin depth & m \\                                       
$\sigma$            & Stefan-Boltzmann constant & (5.67051~10$^{-8}$) W~m$^{-2}$~K$^{-4}$&   $P$                 & Rotation period & s \\                                          
$S_\odot$           & Solar constant at $r$=1 au & (1329) W~m$^{-1}$&                        $\omega$            & = $2\pi / P$ & s$^{-1}$ \\                                      
$r$                 & Distance to the Sun             & au &                                 $\lambda_p$         & ecliptic longitude of the pole & deg \\                         
$\vec{r}$           & Vector to the Sun & m &                                                $\beta_p$           & ecliptic latitude of the pole & deg \\                          
$\Delta$            & Distance to the observer     & au &                                    $\phi_0$            & Initial rotational phase at epoch & deg \\                      
$\epsilon$          & Emissivity & - &                                                       $\alpha$            & (Phase) angle between asteroid-sun-observer & deg \\            
$\eta$              & Beaming parameter & - &                                                $a$                 & Area of a facet & m$^{2}$\\                                     
$\kappa$            & Thermal conductivity & W~m$^{-1}$~K$^{-1}$&                            $S$                 & Shadowing function & \\                                         
$C$                 & Heat capacity & J~kg$^{-1}$~K$^{-1}$&                                  $F_{\vec{\imath},\vec{\jmath}}$ & View factor & \\                                    
$\Gamma$            & Thermal inertia & \tiu&                                                $J_V(\vec{\jmath})$     & Visible radiosity & \\                                      
$\Theta$            & Thermal parameter & - &                                                $J_{IR}(\vec{\jmath})$   & Infrared radiosity & \\                                    
$\rho$              & Material density & kg~m$^{-3}$&                                        $\hat{n}$           & Local normal & m \\                                             
$H$                 & Absolute magnitude of the $H,G$ system &  &                            $\vec{\imath}$      & Vector to the local facet & m \\                                
$G$                 & Slope parameter of the $H,G$ system & &                                $\vec{\jmath}$      & Vector to the remote facet & m \\                               
$V$                 & Actual magntude in the V-band & &                                      $\gamma_C$          & Crater opening angle & deg \\                                   
$D$                 & Diameter & m (or km) &                                                 $\rho_C$ or $f$     & Area density of craters & \\                                    
$A$                 & Bolometric Bond albedo & - &                                           $\phi$              & Emission angle & rad \\                                         
$p_V$               & Geometric visible albedo & - &                                         $f_\lambda(\tau)$   & Infrared flux & W~m$^{-2}$~\micron$^{-1}$ \\                    
$z$                 & Depth in the subsoil & m &                                             $\lambda$           & Wavelength & \micron \\                                         
$t$                 & Time & s \\                                                            $r_p$               & Pore radius of regolith  & m \\

\enddata          
\end{deluxetable} 

% !TEX root = astIVtpm_Rev2.tex
\section{\textbf{THERMOPHYSICAL MODELS}}
\label{S:tpm}

\subsection{Overview}

Different TPMs have been proposed to study the thermal emission of asteroids, comets, planets, and satellites. The first models were motivated by  thermal observations of the lunar surface, which revealed an almost thermally insulating surface that emitted thermal radiation in a non-Lambertian way \citep{Pettit1930ApJ...71...102,Wesselink1948BAN...10...351W}. Heat conduction and radiation scattering models of various rough surfaces were able to reproduce the lunar observations to a good degree 
\citep[e.g.,][]{Smith1967JGR....72.4059S,Buhl1968JGR....73.5281B,Buhl1968JGR....73.7593B,Sexl1971Moon....3..189S,Winter1971Moon....2..279W}, and the derived thermal inertia and surface roughness values matched \emph{in situ} measurements by Apollo astronauts \citep[see][and references therein]{Rozitis2011MNRAS.415.2042R}.  These early lunar models were adapted to general planetary bodies, albeit with an assumed spherical shape, by \cite{Spencer1989Icar...78..337S} and \cite{Spencer1990Icar...83...27S}. The most commonly used asteroid TPMs of \cite{Lagerros1996A&A...310.1011L,Lagerros1997A&A...325.1226L,Lagerros1998A&A...332.1123L}, \citet{Delbo2004PhD}, \citet{Mueller2007PhD}, and \citet{Rozitis2011MNRAS.415.2042R} are all based on \cite{Spencer1989Icar...78..337S} and \cite{Spencer1990Icar...83...27S}. Here we present the basic principles utilized in TPMs; for implementation details, the reader is referred to the quoted works.

\begin{figure}[h]
% \epsscale{1.5}
%\plotone{figs/Baccus_Rough.pdf}
\includegraphics[trim=0 00 00 00,clip,scale=0.38]{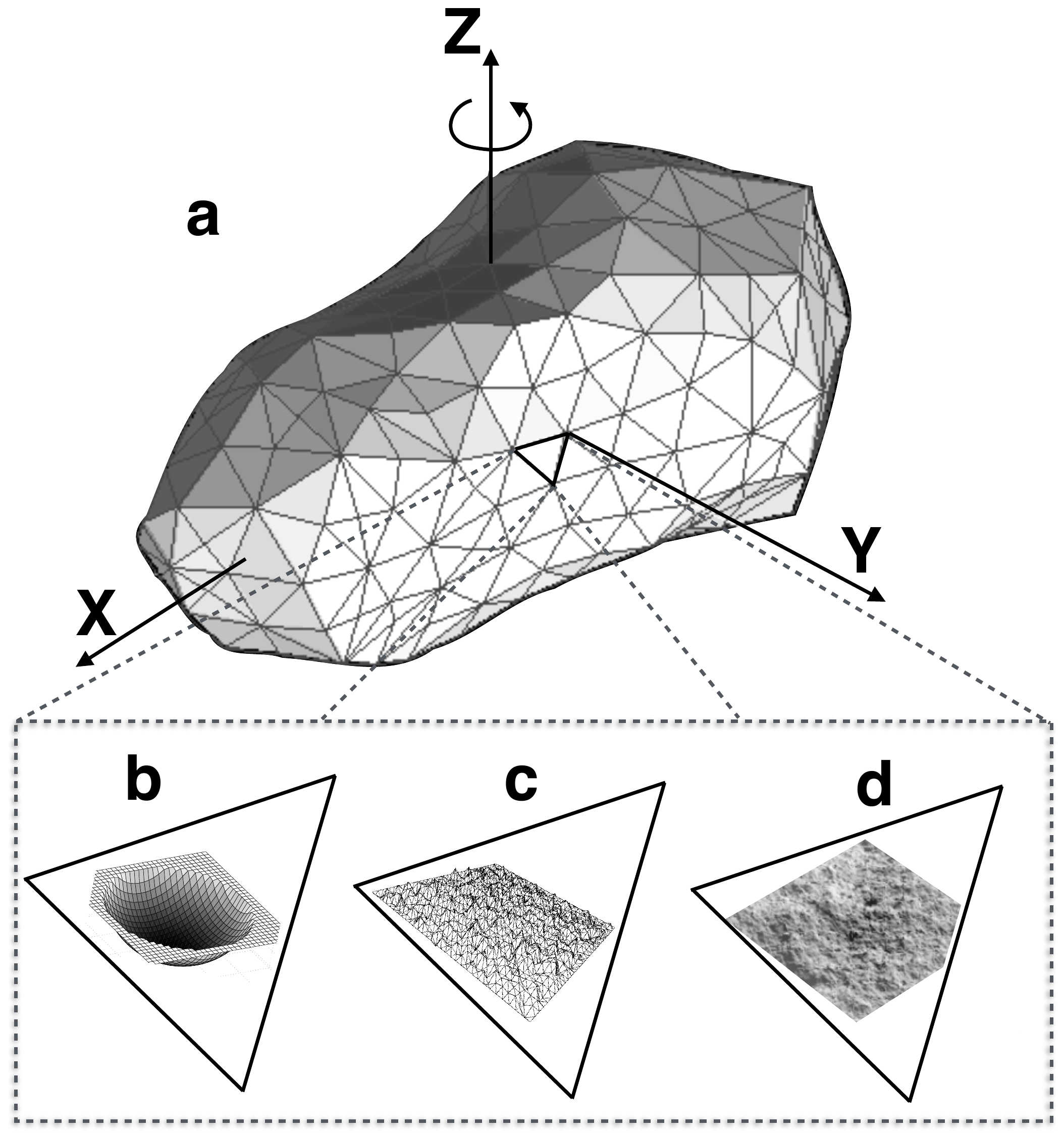}
\caption{\label{F:tpmMesh} \small (a) example of a triangulated 3D shape model as typically used in TPMs (asteroid (2063) Bacchus from http://echo.jpl.nasa.gov/asteroids/shapes/shapes.html). Temperatures are color coded:
white corresponds to the maximum and dark-gray corresponds to minimum temperature. Three different roughness models are sketched in the bottom of the figure: (b) hemispherical section craters; (c) Gaussian surface; (d) fractal surface. Sub-figures b and d are adapted from \citet{Davidsson2015Icarus}, c is from \citet{Rozitis2011MNRAS.415.2042R}.}
% echo bacchus.obj ivar.eph 0.9 200 0.09 0.0 0.0 | runtpm
% glviewmesh -m bacchus.obj -v -w -s 10 -T tmap_temps.dat 20
\end{figure}

All TPMs represent the global asteroid shape as a mesh of (triangular) facets (see Fig.~\ref{F:tpmMesh}) that rotates around a given spin vector with rotation period $P$. In general, utilized shape models are derived from radar observations, inversion of optical light-curves,  in-situ spacecraft images, or stellar occultation timing \citep[see][for a review of asteroid shape modeling]{Durech2014astIV}. If no shape model is available, one typically falls back to   a sphere or an ellipsoid \cite[e.g.,][]{Muller2013A&A...558A..97M,Muller2014PASJ...66...52M,Emery2014Icar..234...17E}. 

The goal is to calculate the thermal emission of each facet of the shape model at a given illumination and observation geometry.  To this end, the temperature of the surface and, in the presence of thermal inertia, the immediate subsurface need to be calculated.
Generally, lateral heat conduction can be neglected as the shape model facets are much larger than the penetration depth of the diurnal heat wave  (i.e., the thermal skin depth), and only 1D heat conduction perpendicular to and into the surface needs to be considered. For  temperature $T$, time $t$, and depth $z$, 1D heat conduction is described by:
\begin{equation}
\rho C \frac{\partial T}{\partial t} = \frac{\partial}{\partial z} \kappa \frac{\partial T}{\partial z}
\label{E:heatDiffusion_TempDependent}
\end{equation}
where $\kappa$ is the thermal conductivity, $\rho$ is the material density, and $C$ is the heat capacity. If $\kappa$ is independent of depth (and, implicitly, temperature independent, see \S\ref{S:tempDependentGamma}), Eq.\ \ref{E:heatDiffusion_TempDependent} reduces to the diffusion equation: 
\begin{equation}
 \frac{\partial T}{\partial t} = \frac{\kappa}{\rho C} \frac{\partial T^2}{\partial^2 z}
%\rho C \frac{\partial T}{\partial t} = 
%       \frac{\partial}{\partial z}
%       \kappa \frac{\partial T}{\partial z} \simeq 
%       \kappa \frac{\partial T^2}{\partial^2 z}
\label{E:heatDiffusion}
\end{equation}
It is useful to define the thermal inertia $\Gamma$ and the thermal skin depth $l_s$
\begin{equation}
\Gamma = \sqrt{\kappa \rho C}
\label{E:Gamma}
\end{equation}
\begin{equation}
l_s = \sqrt{\kappa P / 2 \pi \rho C}. 
\label{E:thermalSkinDepth}
\end{equation}

These material properties are generally assumed to be constant with depth and temperature in asteroid TPMs, but varying properties has been considered in some Moon, Mars, planetary satellites, and asteroids models \citep[e.g][see also section~\ref{S:tempDependentGamma}]{Giese1990Icar...88..372G,Urquhart1997JGR...10210959U,Piqueux2011JGRE..116.7004P,Keihm1984Icar...60..568K,Keim2012Icar..221..395K,Capria2014GeoRL..41.1438C}. 

\begin{figure}[t]
% \epsscale{1.5}
%\plotone{figs/tEquator.pdf}
\includegraphics[trim=40 00 20 00,clip,width=9cm, height=6cm]{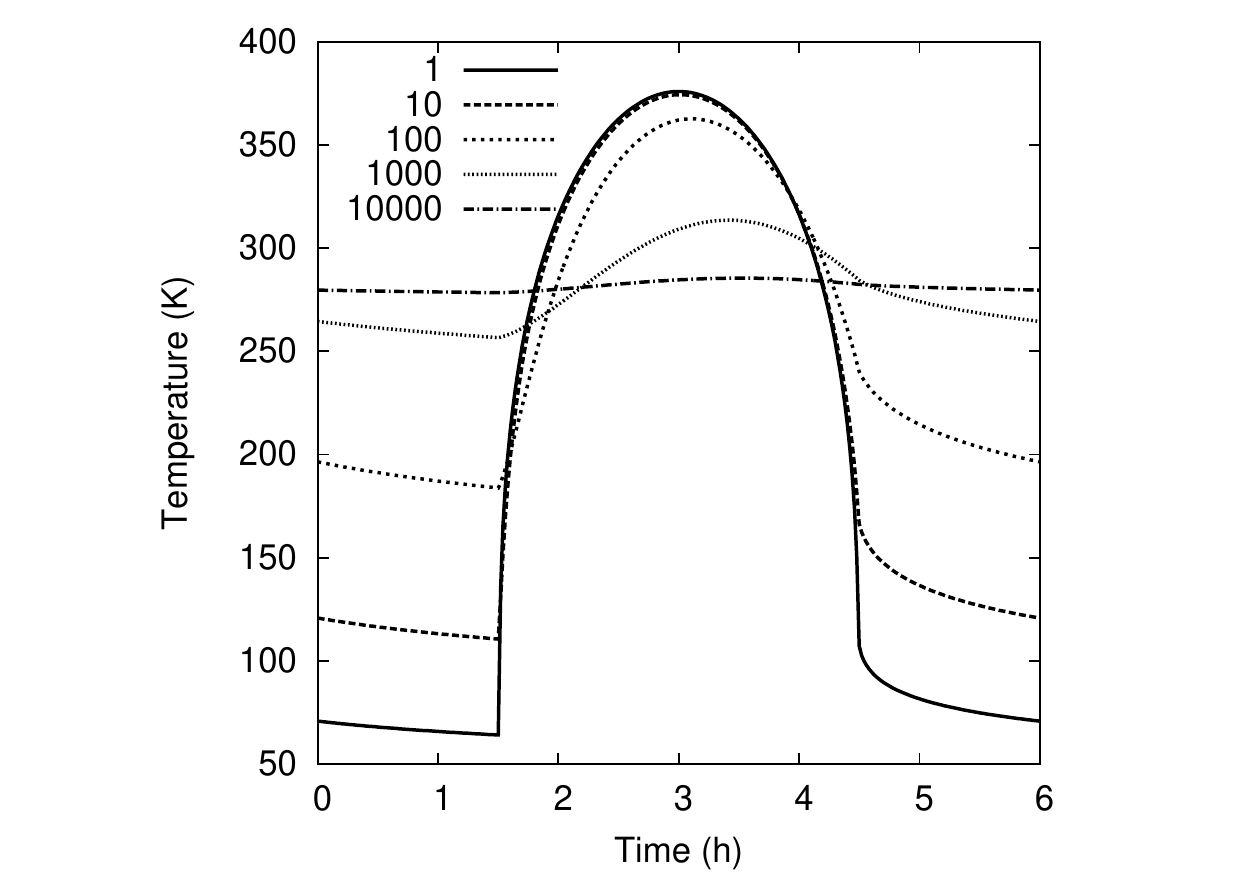}
\caption{ \label{F:diurnalTemperatureCurves} \small Synthetic diurnal temperature curves on the equator of a model asteroid for different values of thermal inertia (in units of \tiu). Increasing thermal inertia smooths temperature contrasts and  causes the temperature peak to occur after the insolation peak at 3 h. The asteroid is situated at a heliocentric distance of r = 1.1 au, has a spin period of 6 h, a Bond albedo of A = 0.1, and its spin axis is perpendicular to the orbital plane. }  
\end{figure}

TPM implementations typically employ dimensionless time and depth variables: $\tau = 2\pi t/P$ and $Z = z/l_s$.  Then, the only remaining free parameter is the dimensionless thermal parameter \citep[$\Theta = \Gamma \sqrt{\omega} / \epsilon \sigma T_{SS}^3$,][]{Spencer1990Icar...83...27S} describing the combined effect of thermal inertia, rotation period, and heat emission into space on the surface temperature distribution (see Fig.~\ref{F:diurnalTemperatureCurves}).

%simplify the equations involved to depend only on the thermal inertia $\Gamma$, and also makes the results scale independent. 
A numerical finite-difference technique is used to solve the 1D heat conduction equation, and an %Newton-Raphson 
iterative technique is used to solve the surface boundary condition. This requires a suitable number of time and depth steps to fully resolve the temperature variations and to ensure model stability (typically, at least 360 time steps and 40 depth steps over 8 thermal skin depths are required). TPMs are run until specified convergence criteria are met (e.g., until temperature variations between successive model iterations are below a specified level) and/or until a specified number of model iterations have been made.

For applications such as the study of the sublimation of water ice from the shallow subsurface of asteroids (e.g., the \emph{Main Belt Comets} or 24 Themis) the heat conduction equation must be coupled with a gas diffusion equation \citep{Schorghofer2008ApJ...682..697S,Capria2012A&A...537A..71C,Prialnik2009MNRAS.399L..79P}. See also \citet{Huebner:2006IssiBook} for a review.

\begin{figure}[t]
% \epsscale{1.5}
\includegraphics[trim=0 00 00 00,clip,scale=0.9]{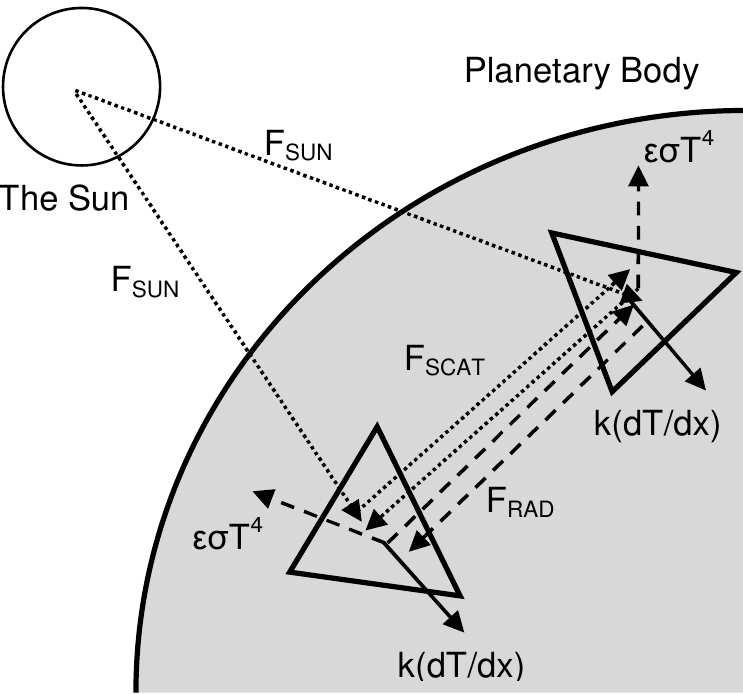}
\caption{\small Diagram illustrating the energy balance and radiation transfer between facets \citep[copied from][]{Rozitis2011MNRAS.415.2042R}. The terms F$_\text{SUN}$, F$_\text{SCAT}$, F$_\text{RAD}$, $k(dT/dx)$ and $\epsilon \sigma T^4$ are the direct sunlight, multiply scattered sunlight, reabsorbed thermal radiation, conducted heat and thermal radiation lost to space, respectively.}
% echo bacchus.obj ivar.eph 0.9 200 0.09 0.0 0.0 | runtpm
% glviewmesh -m bacchus.obj -v -w -s 10 -T tmap_temps.dat 20
\label{F:EnergyBalanceFacets}
\end{figure}

The 1D heat conduction equation is  solved with internal and surface boundary conditions to ensure conservation of energy. Since the amplitude of subsurface temperature variations decreases exponentially with depth, an internal boundary condition is required to give zero temperature gradient at a specified large depth
\begin{equation}
\left ( \frac{\partial T}{\partial z} \right )_{z \gg l_s} = 0.
\end{equation}
A typical surface boundary condition for a facet at point $\vec{\imath}$ with respect to the asteroid origin, at point $\vec{r}$ with respect to the Sun, and with surface normal $\hat{n}$ is then given by
\begin{equation}
\begin{aligned}
& \epsilon \sigma T^4(\vec{\imath},t) - \left( \frac{\partial T(\vec{\imath},t) }{\partial z} \right)_{z=0}  = \\
& \frac{(1-A) S_\odot}{\vec{r}~^3} (\vec{r} \cdot \hat{n}) (1- S(\vec{r},\vec{\imath})) + \\
& (1-A) \int J_V(\vec{\jmath}) F_{\vec{\imath},\vec{\jmath}}~da' + \\
& \epsilon \sigma (1-\epsilon) \int J_{IR}(\vec{\jmath}) F_{\vec{\imath},\vec{\jmath}}~da'
\end{aligned}
\label{E:surfBoundCond}
\end{equation}
%
%where $\epsilon$ is the emissivity (a constant value between 0.9 and 1.0 is typically chosen), and $\sigma$ and $S_\odot$ are the Stefan-Boltzmann and solar constants, respectively. 
The left-hand side of Eq.\ (\ref{E:surfBoundCond}) gives the thermal energy radiated to space and the heat conducted into the subsurface, and the right-hand side gives the input radiation from three different sources: direct solar radiation, multiply scattered solar radiation (i.e., self-illumination), and reabsorbed thermal radiation (i.e., self-irradiation). The two last components are also known as mutual heating (see Fig.~\ref{F:EnergyBalanceFacets}). 

The amount of solar radiation absorbed by a facet depends on the Bond albedo $A$ and any shadows projected on it, which is dictated by $S(\vec{r},\vec{\imath})$ (i.e., $S(\vec{r},\vec{\imath})= 1$ or 0, depending on the presence or absence of a shadow). Projected shadows occur on globally non-convex shapes only, which can be determined by ray-triangle intersection tests of the solar illumination ray \citep[e.g.,][]{Rozitis2011MNRAS.415.2042R} or by local horizon mapping \citep[e.g.,][]{Statler2009Icar..202..502S}. 
%Computationally, it is an $O(N^2)$ problem (where $N$ is the number of facets) to determine which facets block other facets, but for a given geometry they only need to be calculated once and the results can be stored in a lookup table for later use. 
Related to shadowing are the self-heating effects
arising from interfacing facets, which tend to reduce the temperature contrast produced inside concavities. 
The problem here is to determine which facets see other facets, and to calculate the amount of radiation exchanged between them. The former can be determined by ray-triangle intersection tests again, and the latter can be solved using view factors. The view factor $F_{\vec{\imath},\vec{\jmath}}$ is defined as the fraction of the radiative energy leaving the local facet $\vec{\imath}$ that is received by the remote facet $\vec{\jmath}$ assuming Lambertian emission \citep{Lagerros1998A&A...332.1123L}. 
%Like shadowing, these tests only need to be performed once and can be re-used from a lookup table. 
$J_V(\vec{\jmath})$ and $J_{IR}(\vec{\jmath})$ are then the visible and thermal-infrared radiosities of remote facet $\vec{\jmath}$. Either single or multiple scattering can be taken into account, and the latter can be efficiently solved using Gauss-Seidel iterations \citep{Vasavada1999Icar..141..179V}. Most TPMs neglect shadowing and self-heating effects resulting from the global shape for simplicity, but they can be significant on asteroids with large shape concavities \citep[e.g., the south pole of (6489) Golevka:][]{Rozitis2013MNRAS.433..603R}.

\subsection{Modeling asteroid thermal emission}

%Once the surface temperature distribution across a planetary surface has been computed, the emission spectrum at any given phase of rotation can be calculated. When the temperature T(t) at time t for a facet is known, the intensity of radiation it emits $I_{lambda}(t)$ at a desired wavelength $\lambda$ is given by the Planck function:
%\begin{equation}
%\text{plank}
%\end{equation}
%Surface irregularities at scales below the thermal skin depth and comparable to the regolith grain sizes (microscopic roughness) could also affect the overall surface thermal emission and lead to microscopic beaming. See Rozitis 2011.
%Transparency of the soil in the thermal IR. \cite{Hale2002Icar..156..318H}\\
%Emission in the Far IR, sub-mm. 

\begin{figure}[h]
\plotone{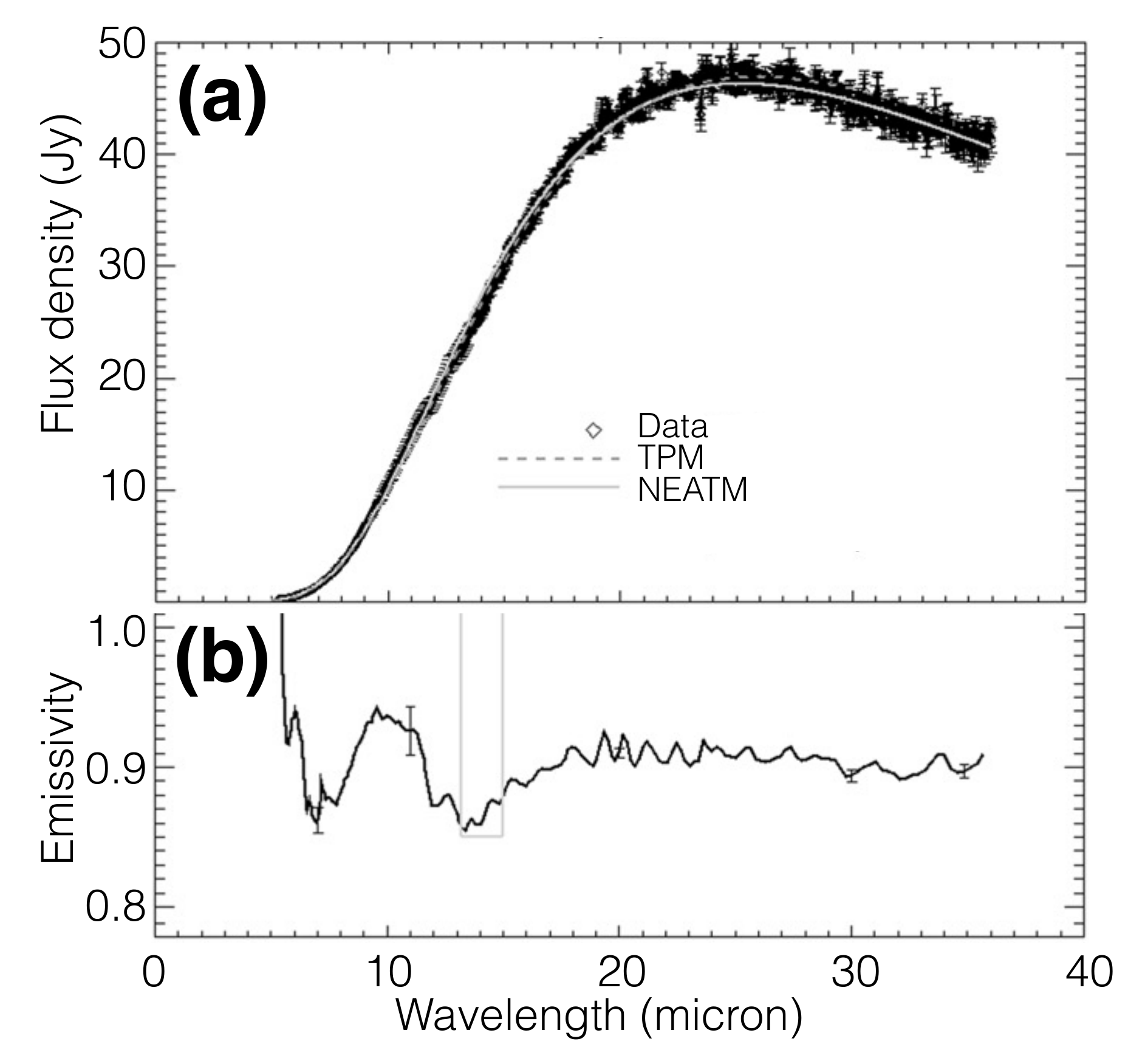}
\caption{\label{F:SED_emissivity} \small (a) Example of SED calculated from a TPM and the NEATM compared to a Spitzer spectrum of (87) Sylvia. (b) 
Spectral emissivity derived from the above: data divided by the NEATM continuum. Figure from \cite{Marchis2012Icar..221.1130M}.}
\end{figure}

\begin{figure}[h]
\plotone{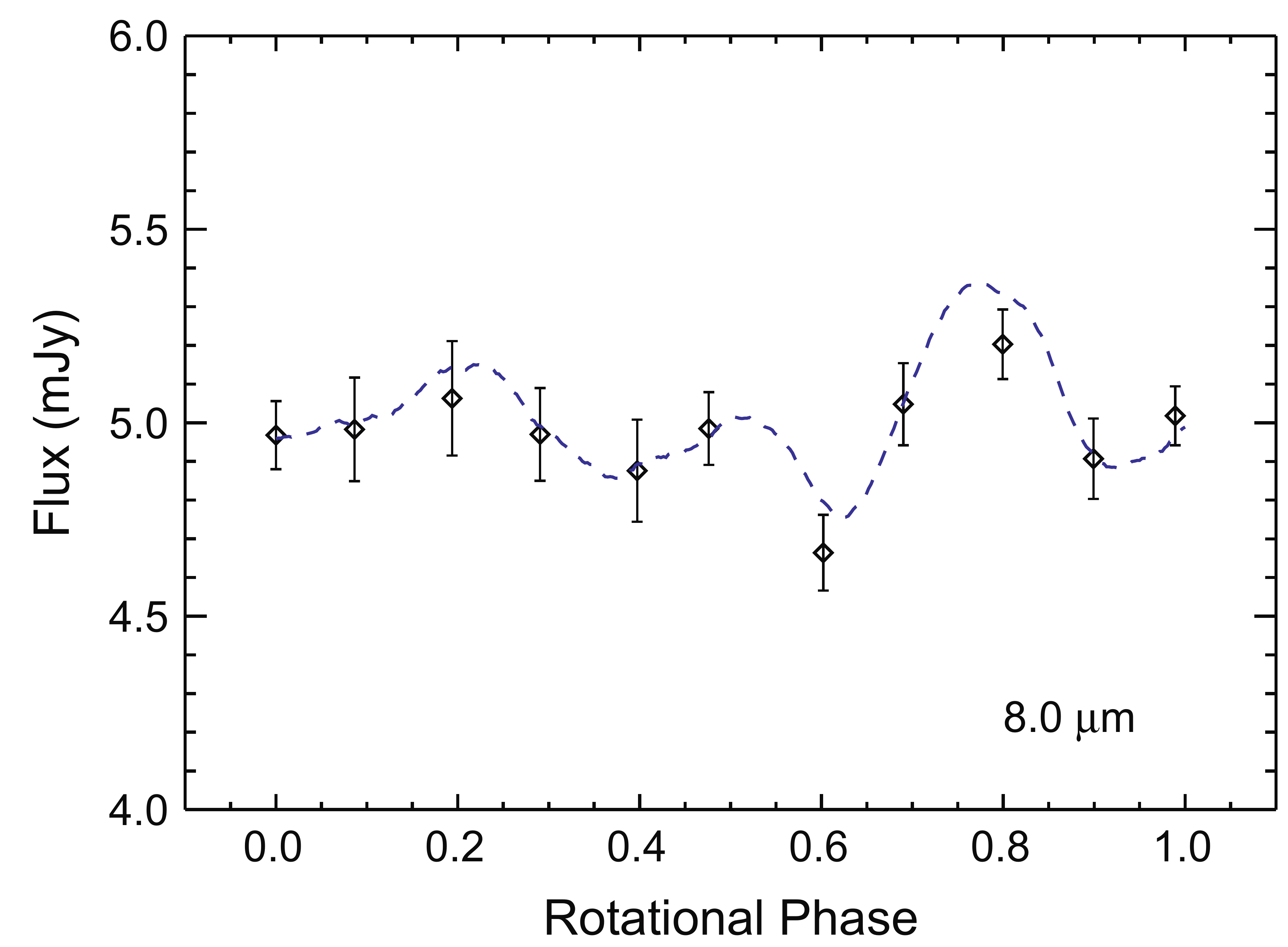}
\caption{\label{F:thermalLC} \small Example of a TPM generated thermal lightcurve (dashed line) and real data for (101955) Bennu. From \cite{Emery2014Icar..234...17E}.}
\end{figure}

Once the surface temperature distribution across an asteroid surface has been computed, the emission spectrum (Fig.~\ref{F:SED_emissivity}) at a given observation geometry and a specified time can be calculated. The monochromatic flux density can also be calculated at wavelengths of interest. When these model fluxes are plotted as a function of the asteroid rotational phase, one obtains the so-called thermal lightcurves (e.g. Fig.~\ref{F:thermalLC}), which can be used to test the fidelity of shape and albedo models typically used as input in the TPM.

When the temperature for a facet is known, the intensity $I_\lambda(\tau)$ at which it emits at wavelength $\lambda$ is given by the Planck function. Assuming Lambertian emission, the spectral flux of the facet seen by an observer is then
\begin{equation}
f_\lambda(\tau) = I_\lambda(\tau) \frac{a}{\Delta^2} \cos \phi
\end{equation}
where $a$ is area of the facet, $\Delta$ is the distance to the observer, and $\phi$ is the emission angle. The total observed flux is obtained by summing the thermal fluxes of all visible shape model facets including any contributions from surface roughness elements contained within them. For disk-integrated measurements, this summation is performed across the entire visible side of the asteroid, whilst for spatially resolved measurements it is summed across facets contained within the detector pixel's field of view.

The assumption of Lambertian emission depends on no directionality induced by surface irregularities at scales below the thermal skin depth. \citet{Davidsson2014Icar..243...58D}  show that surface roughness at sub-thermal-skin-depth scales is quasi-isothermal and is therefore not likely to deviate from Lambertian emission overall. However, radiative transfer processes between the regolith grains could contribute up to 20\% of the observed beaming effects \citep{Hapke1996JGR...10116817H}. \citet{Rozitis2011MNRAS.415.2042R,Rozitis2012MNRAS.423..367R} investigated combined microscopic (regolith grain induced) and macroscopic (surface roughness induced) beaming effects, and demonstrated that the macroscopic effects dominated overall. This was previously found to be the case in directional thermal emission measurements of lava flows on Earth \citep{Jakosky1990GeoRL..17..985J}.

As wavelengths increase to the submillimeter range and above, asteroid regolith becomes increasingly transparent and the observed flux is integrated over increasing depths \citep{Chamberlain2009Icar..202..487C,Keihm2013Icar..226.1086K}.   Modeling such fluxes with typical thermal models (which derive fluxes from surface temperatures, only) requires a significant reduction in effective spectral emissivity.  
For example, 3.2 mm flux measurements of (4) Vesta require an emissivity of $\sim$0.6 to match model predictions \citep{Muller2007A&A...467..737M}.
The reduction in emissivity can be explained by lower subsurface emission temperatures \citep{Lagerros1996A&A...315..625L} and by different subsurface scattering processes dependent on grain size \citep{Redman1992AJ....104..405R,Muller1998A&A...338..340M}.  \citet{Keim2012Icar..221..395K,Keihm2013Icar..226.1086K} attribute the reduced emissivity at submm/mm wavelengths to a higher thermal inertia value of the subsurface layers.
Reduction in emissivity has also been determined at  wavelengths shorter than 4.9 \micron\ for disk-resolved regions of (4) Vesta \citep{Tosi2014Icar..240...36T}.

% !TEX root = astIVtpm_Rev2.tex
\subsection{Surface Roughness}
\label{S:roughness}

Roughness causes an asteroid surface to thermally emit in a non-Lambertian way with a tendency to reradiate the absorbed solar radiation back towards the Sun, an effect known as thermal infrared beaming \citep{Lagerros1998A&A...332.1123L,Rozitis2011MNRAS.415.2042R}. 
It is thought to be the result of two different processes: a rough surface will have elements orientated towards the Sun that become significantly hotter than a flat surface, and multiple scattering of radiation between rough surface elements increases the total amount of solar radiation absorbed by the surface. 
%The surface roughness responsible starts at the thermal skin depth and ranges up to the size of the facets used in an asteroid shape model. 
The relevant size scale ranges from the thermal skin depth to the linear size of the facets in the shape model.
It is included in thermophysical models by typically modeling an areal fractional coverage ($f$) of spherical-section craters (of opening angle $\gamma_C$) within each shape model facet. 
Other more complex forms have been considered, such as Gaussian random \citep{Lagerros1998A&A...332.1123L} or fractal \citep{Groussin2013Icar..222..580G} surfaces or parallel sinusoidal trenches (see sketch of Fig.~\ref{F:tpmMesh}) ,
 but the spherical-section crater produces similar results \citep[in terms of the disk-integrated beaming effect][]{Lagerros1998A&A...332.1123L} and accurately reproduces the directionality of the lunar thermal infrared beaming effect \citep{Rozitis2011MNRAS.415.2042R}. However, it has been shown that the thermal emission depends also on roughness type in addition to roughness level, for disk resolved data \cite{Davidsson2015Icarus}.

%However, it is important to note that, when a particular roughness model is used to interpret observed data, the value of the thermal inertia and degree of roughness calculated from the application of the TPM to thermal infrared data can be model dependent, at least to a certain extent. \citet{TBD Davidsson_2014} modeled the  thermal emission using different roughness models, as a function of illumination and viewing geometries, degrees of surface roughness and wavelengths, and found that the thermal emission can dependent on the roughness model for certain illumination and viewing geometries, even when the degrees of roughness (e.g. same value of $\bar \theta$; see Eq.~\ref{E:thetaBar}) are identical. It is therefore important to mention the roughness model used in the TPM for thermal inertia determination. 
Spherical-section craters are typically implemented, as the required shadowing and view-factor calculations can be performed analytically \citep{Emery1998Icar..136..104E,Lagerros1998A&A...332.1123L}. Heat conduction can be included by dividing the crater into several tens of surface elements where the same equations listed above can be applied. Alternatively, the temperature distribution within the crater resulting from heat conduction, $T_\text{crater}(\Gamma)$, can be approximated using
\begin{equation}
\frac{T_\text{rough}(\Gamma)}{T_\text{rough}(0)} = \frac{T_\text{smooth}(\Gamma)}{T_\text{smooth}(0)} 
\label{E:LagerrosApprox}
\end{equation}
where $T_\text{rough}(0)$ can be calculated analytically assuming instantaneous equilibrium \citep{Lagerros1998A&A...332.1123L}. 
$T_\text{smooth}(0)$ and $T_\text{smooth}(\Gamma)$  are the corresponding smooth-surface temperatures, which can be calculated exactly. This approximation is computationally much cheaper  than the full implementation.  However, it does not work on the night side of the asteroid and temperature ratios diverge near the terminator \citep{Mueller2007PhD}. An even simpler alternative is to multiply the smooth-surface temperatures by a NEATM-like $\eta$ value \citep[e.g.][]{Groussin2011A&A...529A..73G}. Whilst this alternative might produce the correct disk-integrated color temperature of the asteroid, it does not reproduce the directionality of the beaming effect. Indeed, roughness models predict a limb-brightening effect \citep{Rozitis2011MNRAS.415.2042R}, which is seen in spatially-resolved measurements of (21) Lutetia by Rosetta \citep{Keim2012Icar..221..395K}.

The above implementations neglect lateral heat conduction,
although the spatial scales representing surface roughness can, in some cases, become comparable to the thermal skin depth.
%\citet{Davidsson2015Icarus} have investigated this approximation using a detailed 3D heat conduction model for arbitrary rough surfaces, and find deviations occur for predominantly positive relief topography at sub thermal skin depth scales (e.g. isolated rocks or extreme Gaussian random surfaces). 
Modeling of 3D heat conduction inside rocks the size of the thermal skin depth  has demonstrated that their western sides (for a prograde rotator; eastern sides for a retrograde rotator) are generally warmer than their eastern sides, which could result in a tangential-YORP effect that predominantly spins up asteroids \citep{Golubov2012ApJ...752L..11G}. Other than this, it appears that the 1D heat conduction approximation still produces satisfactory results.
	
In thermophysical models, the degree of surface roughness can be quantified in terms of the Hapke mean surface slope
\begin{equation}
\tan \bar{\theta} = \frac{2}{\pi} \int_0^{\pi/2} a(\theta) \tan \theta~ d\theta
\label{E:thetaBar}
\end{equation}
where $\theta$ is the angle of a given facet from horizontal, and $a(\theta)$ is the distribution of surface slopes \citep{Hapke1984Icar...59...41H}. Alternatively, it can be measured in terms of the RMS surface slope \citep{Spencer1990Icar...83...27S}. This then allows comparison of results derived using different surface roughness representations \citep[e.g., craters of different opening angles and fractional coverages, or different Gaussian random surfaces:][]{Davidsson2015Icarus}, and comparison against roughness measured by other means. It has been demonstrated that different roughness representations produce similar degrees of thermal infrared beaming when they have the same degree of roughness measured in terms of these values \citep{Spencer1990Icar...83...27S,Emery1998Icar..136..104E,Lagerros1998A&A...332.1123L,Rozitis2011MNRAS.415.2042R}. %{\color{red} TBD however, see the results of \citet{Davidsson2015Icarus} }

% !TEX root = astIVtpm_Rev2.tex
%\section{IMPLEMENTATION OF TPMs. \\ How models are used to analyze data}
\section{DATA ANALYSIS USING A TPM}
\label{S:implementationTPM}
%\section{Implementation of TPMs:  how TPMs are used to analyze data}
\subsection{Thermal infrared spectro-photometry}
% \cc{Ben, please start write this; Marco will add things\\
% describe the fit procedure to disk integrated \\
% what is the effect on the thermal inertia determination ? is it within the error bars ? \\
% effect of shape on thermal inertia determination (Hanus, Rozitis, Emery) \\
% color correction for WISE and IRAS and other broadband filters. ??? \\}

Physical properties than can be derived from TPM fits to disk-integrated thermal observations include the diameter, geometric albedo,  thermal inertia and roughness.
In practically all cases, the absolute visual magnitude $H$ is known, establishing a link between $D$ and $p_V$ and reducing the number of TPM fit parameters by one:
\begin{equation}
D(\text{km})=1329~p_V^{-1/2}~10^{-H/5},
\end{equation}
\citep{Fowler:1992wd,Vilenius2012A&A...541A..94V}. 
Frequently, the rotational phase during the thermal observations is not sufficiently well known and has to be fitted to the thermal data
\cite[e.g.,][]{Harris2005Icar..179...95H,Ali-Lagoa2014A&A...561A..45A}. 
%If accurate rotational phasing information is available (e.g., from extensive light-curve or radar observations) then this can be kept fixed at the appropriate value. 
In some cases, TPMs can be used to constrain the orientation of the spin vector of an asteroid, with $\lambda_p$ and $\beta_p$ treated as free parameters \citep[as demonstrated e.g., by][note that in the case of 101955 Bennu the radar-constrained pole solution was not yet known]{Muller2013A&A...558A..97M,Muller2012A&A...548A..36M}. 
Moreover, \cite{Muller2014A&A...566A..22M} successfully performed  a TPM analysis of an asteroid (99942 Apophis) in a non-principal axis rotation state for the first time.  

\begin{figure}[h]
\plotone{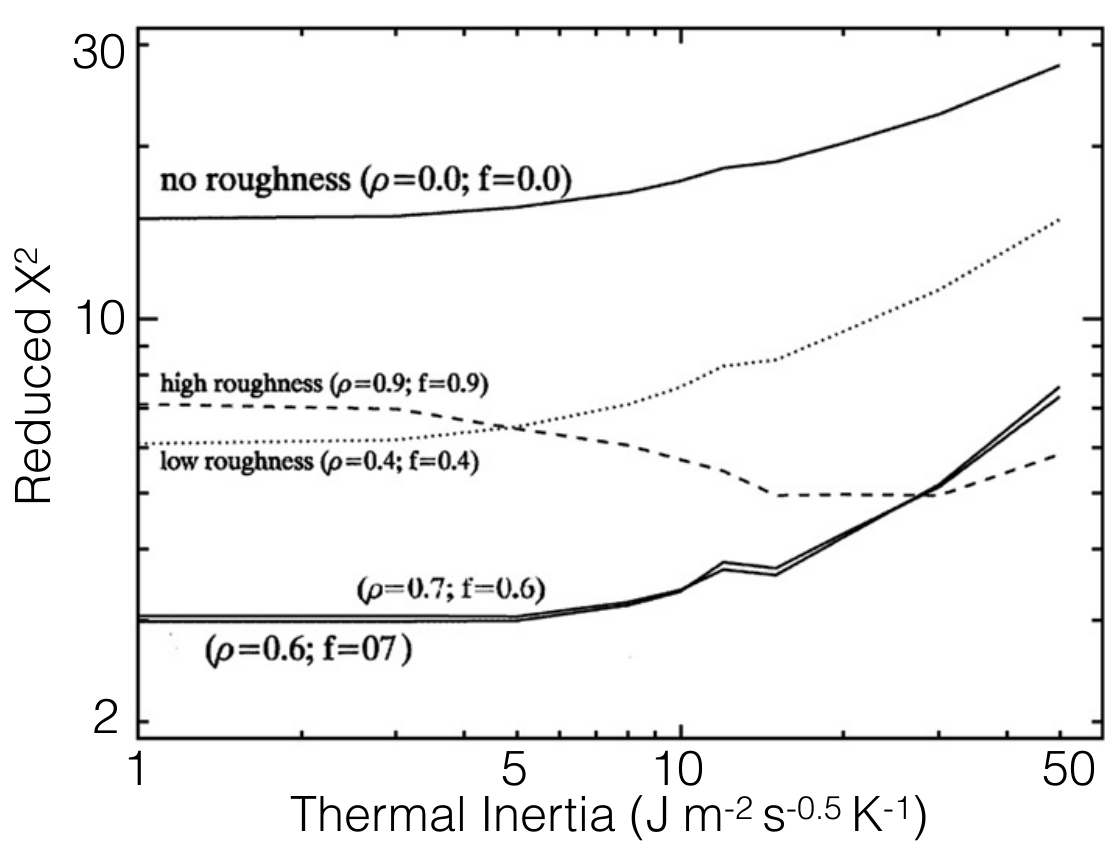}
 \caption{\label{F:roughness_TLC}  (21) Lutetia: a TPM fit that allows surface roughness to be constrained. The quantity $\rho$ here is not the bulk density of the body, but it is the r.m.s. of the slopes on the surface. It is related to the ratio between the diameter and the depth of spherical section craters \citep{Lagerros1998A&A...332.1123L} in this particular case. $f$ is the areal fraction of each facet covered with craters. From \cite{ORourke2012P&SS...66..192O}.}
\end{figure}

The thermal effects of thermal inertia and surface roughness are difficult to tell apart.  
A commonly used approach is to use four different roughness models
corresponding to no, low, medium, and high roughness, with each model leading to a different thermal-inertia fit \citep{Mueller2007PhD,Delbo2009P&SS...57..259D}; frequently, the scatter between these four solutions accounts for the bulk of the uncertainty in  thermal inertia.
However, in some lucky cases, data do allow the effects of roughness and thermal inertia to be disentangled.  This requires good wavelength coverage straddling the thermal emission peak and good coverage in solar phase angle, such that both the morning and afternoon sides of the asteroid are seen.  See Fig.~\ref{F:roughness_TLC} for an illustration.

The best-fitting model parameters are those that minimize $\chi^2$.  Their uncertainty range is spanned by the values that lead to $\chi^2$ within a specified threshold of the best fit, depending on the number of free fit parameters.
Ideally, the reduced $\chi^2$ of the best fit should be around unity.
 However, due to systematic uncertainties introduced in thermal infrared observations (e.g., flux offsets between different instruments), and/or in the thermophysical modeling, it is not uncommon to get large reduced $\chi^2$ values. Large $\chi^2$ values are also obtained when the assumed shape model differs significantly from the asteroid's true shape \citep{Rozitis2014A&A...568A..43R}. In particular, if the spatial extent of the shape model's z-axis is wrong, this can lead to diameter determinations that are inconsistent with radar observations 
\citep[e.g., for 2002 NY$_{40}$ and (308635) 2005 YU$_{55}$ in][respectively]{Muller2004A&A...424.1075M,Muller2013A&A...558A..97M}, and/or two different thermal inertia determinations \citep[e.g., the two different results produced for (101955) Bennu by][]{Emery2014Icar..234...17E,Muller2012A&A...548A..36M}.
In some works, the asteroid shape model has also been optimized during the thermophysical fitting to resolve inconsistencies with radar observations \citep[e.g., (1862) Apollo and (1620) Geographos in][respectively]{Rozitis2013A&A...555A..20R,Rozitis2014A&A...568A..43R}.

We remind the reader that the accuracy of the physical properties (in particular the value of $\Gamma$) of asteroids derived from TPM depends on the quality of the thermal infrared data, coverage in wavelength, phase, rotational, and aspect angle. The accuracy of the shape model and of the $H$ and $G$ values are also important \citep[see e.g.][]{Rozitis2014A&A...568A..43R}. The derived thermal inertia value often depends on the assumed degree of roughness and it is usually affected by large errors (e.g. 50 or 100 \%, see Tab.~\ref{T:TI}). Care must be used in accepting TPM solutions purely based on the goodness of fit (e.g. the value of the $\chi^2$), as they can be dominated by one or few measurements with unreliable small errors or calibration offsets between measurements from different sources. 

\subsection{Thermal Inertia and Thermal Conductivity}
\label{S:thermalInertia} 

As asteroids rotate, the day-night cycle causes cyclic temperature variations that are controlled by the thermal inertia (defined by Eq. \ref{E:Gamma}) of the soil and the rotation rate of the body. 
In the limit of vanishing thermal inertia, the surface temperature would be in instantaneous equilibrium with the incoming flux, depending only on the solar incidence angle (as long as self heating can be neglected); surface temperatures would peak at local noon and would be zero at night.
In reality, thermal conduction into and from the subsoil %(lateral heat conduction can typically be neglected) 
causes a certain thermal memory, referred to as \emph{thermal inertia}.  This smoothens the diurnal temperature profile, leads to non-zero night-side temperatures, and causes the surface temperature to peak on the afternoon side, as shown in Fig.~\ref{F:diurnalTemperatureCurves}, thereby causing the Yarkovsky effect (\S~\ref{S:YarkoYORP}).

%\begin{figure}[t]
%% \epsscale{1.5}
%%\plotone{figs/tEquator.pdf}
%\includegraphics[trim=40 00 20 00,clip,scale=0.9]{figs/tEquator.pdf}
%
%\caption{ \label{F:diurnalTemperatureCurves} \small Synthetic diurnal temperature curves on the equator of a model asteroid for different values of thermal inertia (in units of \tiu). Increasing thermal inertia smooths temperature contrasts and additionally causes the temperature peak to occur after the insolation peak at 3 h. The asteroid is situated at a heliocentric distance of r = 1.1 AU, has a spin period of 6 h, a Bond albedo of A = 0.1, and its spin axis is perpendicular to the orbital plane. }  
%\end{figure}

The mass density $\rho$, the specific heat capacity $c$, and the thermal conductivity $\kappa$, and correspondingly $\Gamma$ itself, must be thought of as effective values representative of the depth range sampled by the heat wave, which is typically in the few-\centi\metre\ range.  In turn, thermal-inertia values inform us about the physical properties of the top few centimeters of the surface, not about bulk properties of the object.

As will be discussed below (\S\ \ref{S:tiLabData}), 
$\rho$ and $c$ of an asteroid surface can plausibly vary within a factor of several, % around $\rho\sim\unit{1,000}{\kilo\gram\per\metre\cubed}$, $c\sim\unit{1,000}{\joule\per\kilo\gram\per\kelvin}$ 
while plausible values of $\kappa$ span a range of more than 4 orders of magnitude.  It is therefore not unjustified to convert from $\Gamma$ to $\kappa$ and back using reference values for $\rho$ and $c$ (note that  Yarkovsky/YORP models tend to phrase the thermal-conduction problem in units of $\kappa$, while TPMs tend to be formulated in units of $\Gamma$, which is the observable quantity).

Importantly, the $\kappa$ of finely powdered lunar regolith is 3 orders of magnitude lower than that of compact rock (compact metal is even more conductive by another order of magnitude).
This is because radiative thermal conduction between regolith grains is significantly less efficient than phononic heat transfer within a grain.  A fine regolith, an aggregate of very small grains, is a poor thermal conductor and displays a low $\Gamma$.
Thermal inertia 
can therefore be used to infer the presence or absence of thermally insulating powdered surface material.
In extension, thermal inertia can be used as a proxy of regolith grain size.
The required calibration under Mars conditions (where the tenuous atmosphere enhances thermal conduction within pores compared to a vacuum) was obtained by \cite{Presley1997JGR...102.6535P,Presley1997JGR...102.6551P} and used in the analysis of thermal-inertia maps of Mars \citep[see][]{Mellon2000Icar..148..437M,Putzig2007Icar..191...68P}.
%% checked Putzig et al., 2014: no update on Mars modeling since 2007 -- see also http://www.boulder.swri.edu/inertia/
Similar progress in asteroid science was slowed down by the lack of corresponding laboratory measurements under vacuum conditions 
(but see below for recent lab measurements of meteorites).  However, \citet{Gundlach2013Icar..223..479G} provided a calibration relation based on heat-transfer modeling in a granular medium.
%%%% This is mentioned later, doesn't need to go here (I guess)
%%%% MM, 2015/02/16
%Basing themselves on a subset of the thermal-inertia data presented in Table\ \ref{T:TI},
%they find a typical regolith grain size in the \milli\metre--\centi\metre\ range for asteroids up to a diameter of $\sim\unit{100}{\kilo\metre}$, while larger objects show much finer regolith in the 10--\unit{100}{\micro\metre} range.

%Information about the typical regolith grain size is important for studies of surface age, with older surfaces expected to display finer regolith.
%Classical models explain asteroidal regolith through the 
%break-up of boulders by micrometeorites \citep{Horz1997M&PS...32..179H}.
%More recently, thermal fatigues was invoked to grind coarse regolith into finer regolith
%\citep[see \S \ref{S:thermalCracking} and][]{Delbo2014Natur.508..233D}.
%YORP-induced shape changes, possibly leading even to binary formation, could bury fine regolith, however, and lead to surface rejuvenation \citep[and references therein]{Delbo2011Icar..212..138D}.

\subsection{Temperature dependence of thermal inertia}
\label{S:tempDependentGamma}
Thermal inertia is a function of temperature \citep{Keihm1984Icar...60..568K}, chiefly because the thermal conductivity is.
In general, for a lunar-like regolith the thermal conductivity is given by:
\begin{equation}
\kappa = \kappa_b + 4 \sigma r_p T^3
\label{E:kappa}
\end{equation}
where $\kappa_b$ is the solid-state thermal conductivity (heat conduction by phonons) and $r_p$ is the radius of the pores of the regolith. The term proportional to $T^3$ is due to the heat conduction by photons.  
Equation~\ref{E:kappa} is often written in the form $\kappa = \kappa_0(1+\chi T^3)$ \citep[e.g.,][]{Vasavada1999Icar..141..179V}. Note that $\kappa_b$ is itself a function of $T$ \citep{Opeil2010Icar..208..449O}. There is extensive literature on the $T$-dependence of the conductivity of lunar regolith \citep[e.g.,][and references therein]{Vasavada1999Icar..141..179V}. A theoretical description of the temperature dependence of $\kappa$ in regoliths is given by \citet{Gundlach2013Icar..223..479G}.

\begin{figure}[h]
\plotone{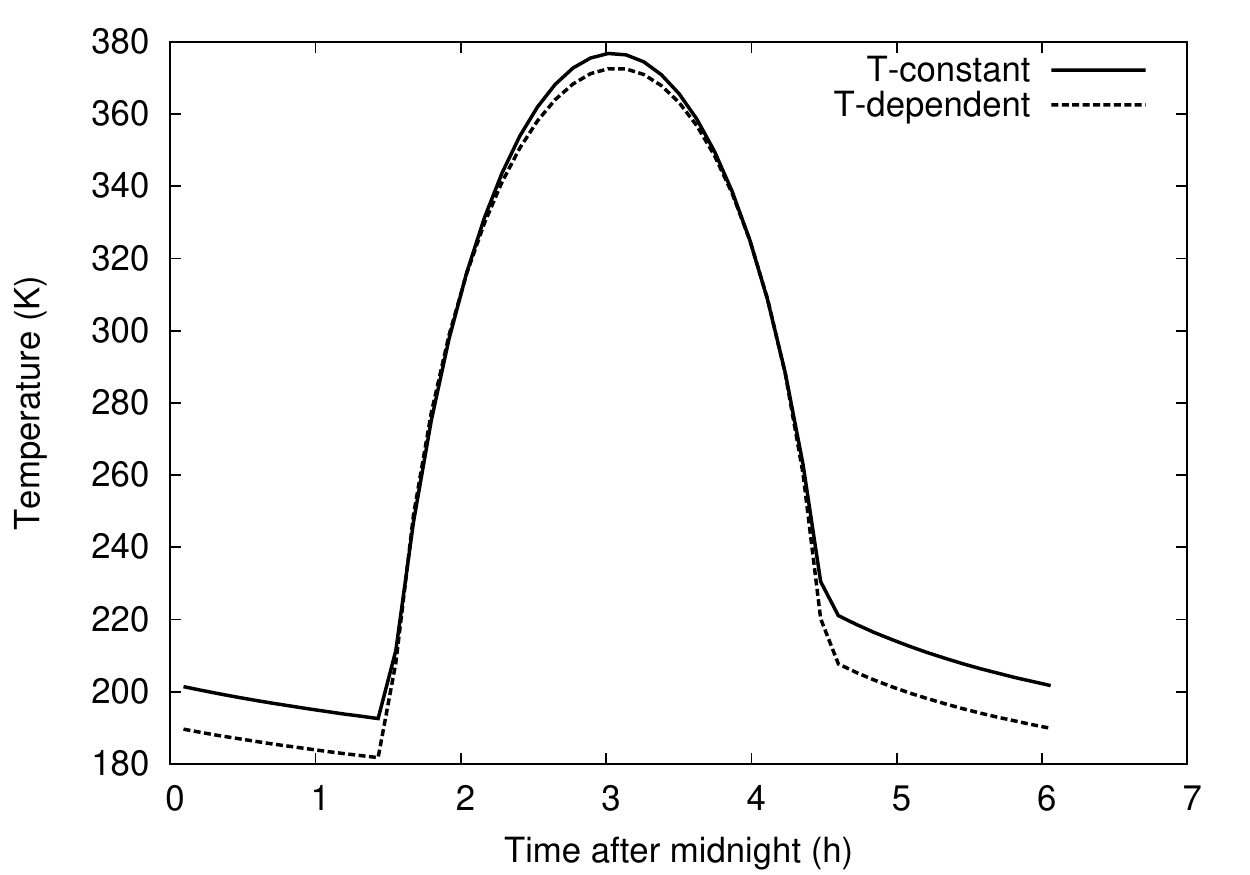}
\caption{\label{F:diurnalCurve2} \small Diurnal temperature curves at the equator of an asteroid with $A$=0.1, $P$=6h, $\epsilon$=1.0 at a heliocentric distance of 1.0 au.
Solid curve: constant thermal conductivity $\kappa$=0.02 \Ku.
Dashed-curved: temperature-dependent heat conductivity $\kappa =  10^{-2}(1+0.5 (T/250)^3)$ \Ku.}
\end{figure}

Asteroid TPMs typically neglect the temperature dependence of $\Gamma$.  This is uncritical for typical remote observations, which are dominated by the warm sunlit hemisphere
\citep[see Fig.~\ref{F:diurnalCurve2} and][]{Capria2014GeoRL..41.1438C,Vasavada2012JGRE..117.0H18V}.
In the analysis of highly spatially resolved observations, however, the temperature dependence must be considered, certainly when analyzing night-time observations on low-$\Gamma$ asteroids.
% (note that the heat capacity is also temperature dependent, but its contribution to the temperature dependence of thermal inertia is less important).  %% is implied in the first sentence of the section
Note that for temperature-dependent $\kappa$, Eq.~\ref{E:heatDiffusion_TempDependent} must be used instead of Eq.~\ref{E:heatDiffusion} \citep[see, e.g.,][]{Capria2014GeoRL..41.1438C}.
%\begin{equation}
%\rho C \frac{\partial T}{\partial t} = \frac{\partial}{\partial z} \kappa \frac{\partial T}{\partial z}
%\end{equation}

% k= kL + kR ; kR = 4 sigma rp T3  ; kL = (A + B T)-1 see table pag 73 Maurin PhD;
%CV(T) see eq. 4.34 Maurin PhD; ACM poster of Consolmagno; 

Caution must be exercised when comparing thermal-inertia results obtained at different heliocentric distances $r$, i.e., at different temperatures.  All other things being equal, $T^4\propto r^{-2}$.  Assuming that the $T^3$ term dominates in Eq.\ \ref{E:kappa}, the thermal inertia of a test object scales with \citep[see also][]{Mueller2010Icar..205..505M}:
\begin{equation}
\Gamma\propto\sqrt{\kappa}\propto T^{3/2}\propto r^{-3/4}.
\label{E:Gamma_r3/4}
\end{equation}

\subsection{Binary Asteroid TPM}
\label{sect:binaryTPM}

A rather direct determination of thermal inertia can be obtained by observing the thermal response to eclipses and their aftermath, allowing one to see temperature changes in real time.  Such observations have been carried out for planetary satellites such as the Galilean satellites \citep{Morrison1973Icar...18..224M}, and our Moon \citep{Pettit1940ApJ....91..408P,Shorthill1973Moon....7...22S,Lawson:2003uv,Lucey2000SPIE.4091..216L,Fountain1976Moon...15..421F}. \citet{Mueller2010Icar..205..505M} report the first thermal observations of eclipse events in a binary Trojan asteroid system, (617) Patroclus, where one component casts shadow on the other while not blocking the line of sight toward the observer.  
%The system couldn't be spatially resolved at thermal wavelengths, therefore the  change in system-integrated flux is used to monitor the eclipse-induced cooling and later warming up in real time.  Such highly time-constrained, differential observations are challenging but not impossible to do using space-based facilities such as Spitzer.  

The thermophysical modeling of eclipse events is relatively straightforward, assuming the system is in a tidally locked rotation state typical of evolved binary systems.  In that case, the components' spin rates match one another, and their spin axes are aligned with that of the mutual orbit.  The system is therefore at rest in a co-rotating frame and can be modeled like a single object with a non-convex (disjoint!) global shape.  Eclipse effects are fully captured, provided that shadowing between facets is accounted for.  The two hemispheres that face one another can, in principle, exchange heat radiatively.  
This is negligible for typical binary systems, however.

As discussed above, the thermal effects of roughness and thermal inertia can be hard to disentangle.
%In the classical application of TPMs, one uses the thermal-infrared data to constrain the diurnal surface temperature distribution on the asteroid, which is affected by both the thermal inertia and the roughness. 
In the case of eclipse measurements, which happen at essentially constant solar phase angle,  the effect of surface roughness is much less of a confounding factor. This is because the variation in the thermal signal is dominated by the temperature change induced by the passing shadow, which is a strong function of thermal inertia.
%temperature rate of change, which is controlled by the value of the thermal inertia, in response to the passing shadow.  
%In turn, large systematic uncertainties can be induced by uncertainties in the orbit of the mutual orbit and/or the size ratio between the two components, which therefore need to be known very accurately for the analysis of the thermal infrared measurement.  In practice, this is a lesser concern because such observations can only be planned if eclipse events can be predicted ahead of time, i.e., the mutual orbit must be known rather accurately to begin with.

It must be kept in mind that the duration of an eclipse event is short compared to the rotation period.  The eclipse-induced heat wave therefore probes the subsoil less deeply than the diurnal heat wave does (the typical heat penetration depth is given by Eq. \ref{E:thermalSkinDepth} with $P$ equal to the duration of the eclipse event). 
A depth dependence of thermal inertia (see \S~\ref{S:tiDepth}) could manifest itself in different thermal-inertia determinations using the two different measurement methods. 
%As discussed in \citet{Mueller2010Icar..205..505M}, such observations have been made on planetary satellites, but not on asteroids.

\subsection{Thermal-infrared interferometry}
Interferometric observations of asteroids in the thermal infrared measure the spatial distribution of the thermal emission along different directions on the plane of sky, thereby constraining the distribution in surface temperature and hence thermal inertia and roughness.
Provided the asteroid shape is known, interferometry can be used to break the aforementioned degeneracy between thermal inertia and roughness from a single-epoch observation
\citep{Matter2011Icar..215...47M,Matter2013Icar..226..419M}.
Interferometry also allows a precise determination of the size of an asteroid \citep{Delbo2009ApJ...694.1228D}. 

Spatial resolutions between 20 and 200 milli-arcseconds can be obtained from the ground \citep[see][for a review and future perspectives of the application of this technique]{Durech2014astIV}. 

%Since the publication of \emph{Asteroids III}, the Mid-Infrared Interferometric Instrument (MIDI) of the the Very Large Telescope Interferometer (VLTI) of the European Southern Observatory (ESO) has been used to obtain measurements of asteroids between 8 and 13 \micron\ \citep{Delbo2009ApJ...694.1228D,Matter2011Icar..215...47M,Matter2013Icar..226..419M}. 
While for the determination of asteroid sizes and shapes from interferometric observations in the thermal infrared, simple thermal models can be used \citep{Delbo2009ApJ...694.1228D,Carry2015Icar..248..516C}, a TPM was utilized to calculate interferometric visibilities of asteroids in the thermal infrared for the observations of (41) Daphne \citep{Matter2011Icar..215...47M} and (16) Psyche \citep{Matter2013Icar..226..419M}. 

Mid-infrared interferometric instruments measure the total flux and the visibility of a source, the latter being related to the intensity of the Fourier Transform (FT) of the spatial flux distribution along the interferometer's baseline projected on the plane of  sky. Thus, the data analysis procedure consisted in generating images of the thermal infrared emission of the asteroids at different wavelengths as viewed by the interferometer and then in obtaining the model visibility and flux for each image. The former is related to the FT of the image, the latter is simply the sum of the pixels. The free parameters of the TPM (size, thermal inertia value and roughness) are adjusted in order to minimize the distance between the disk integrated flux and visibility of the model, and the corresponding observed quantities
\citep[see][for further information]{Matter2011Icar..215...47M,Matter2013Icar..226..419M,Durech2014astIV}. Some results from these observational programs are discussed in \S~\ref{S:results} and in the chapter by \cite{Durech2014astIV}.

%%%%%%%%% DISK RESOLVED %%%%%%%%%%%%
\subsection{Disk-resolved data and retrieval of temperatures}
The availability of disk-resolved thermal-infrared observation has been increasingly  steadily over  the  years: the ESA mission \emph{Rosetta} 
%launched in 2004, on its way to the comet 67P/Churyumov-Gerasimenko, 
performed two successful flybys to the asteroids (2867) Steins (in 2008) and (21) Lutetia (in 2010) \citep{Barucci2014astIV}. In 2011, the NASA mission \emph{Dawn} began its one-year orbiting of (4) Vesta \citep{Russell2012Sci...336..684R}; Dawn is reaching (1) Ceres at the time of writing.  
JAXA's \emph{Hayabusa II} sample-return mission will map its target asteroid (162173) 1999 JU$_3$ using the thermal infrared imager (TIR) aboard the spacecraft \citep{Okada2013LPI....44.1954O}; NASA's 
\emph{OSIRIS-REx} mission and its thermal spectrometer OTES will do likewise for its target asteroid (101955) Bennu (in 2018-2019).
These data can be used to derive surface-temperature maps, from which maps of
 thermal inertia and roughness can be derived. 

Three different methods are used to measure surface temperatures from orbiting spacecraft: 
bolometry, mid-infrared spectroscopy, and near-infrared spectroscopy. 
In the following, we will elaborate on the challenges posed by these different methods, and on their dependence on
spectral features, surface roughness, illumination geometry, and viewing geometry.

Bolometers measure thermal flux within a broad bandpass in the infrared, approximating the integral of the Planck function, $U = \sigma T_e^4$ \citep[e.g.,][]{Kieffer1977JGR....82.4249K,Paige2010SSRv..150..125P}. The temperature derived in this way (\textbf{effective temperature}) is directly relevant to the energy balance on the surface. Since the bolometric flux is spectrally integrated, the resulting temperature is fairly insensitive to spectral emissivity variations, as long as the bolometric emissivity (weighted spectrally averaged emissivity) is known or can be reasonably approximated. 

Temperatures derived from mid-infrared spectrometry, on the other hand, are typically \textbf{brightness temperatures,} i.e., the temperature of a black body emitting at the wavelength in question. It is generally assumed that at some wavelength, the spectral emissivity is very close to 1.0, and the brightness temperature at this wavelength is taken as the surface temperature.  %Because of the strong Si-O vibrational stretching modes near 10 \micron\ in silicates, this assumption is generally valid for silicate-rich asteroid observations, as long as the spectrometer includes the 7 to 13-\micron\ spectral range.  One potential complication for asteroids is that scattering can change from surface-dominated to volume-dominated as the grain size decreases and the regolith becomes more porous (or fluffy). In this case, the wavelength at which the maximum emissivity occurs can change dramatically \citep[e.g.,][]{Logan1973Icar...18..451L,Emery2006Icar..182..496E,2008Natur.454..858V}. The interpretation of some of the extremely low thermal inertias mentioned later (\S~\ref{S:results}) is that many asteroids have such fine-grained, porous regoliths.  Spectrometers offer the obvious benefit of also returning information on the spectral emissivity, which can be used to interpret surface compositions.

Spacecraft sent to asteroids (and/or comets) have more commonly been instrumented with near-infrared spectrometers (e.g., $\lambda <$ 5 \micron) rather than mid-infrared spectrometers.  The long-wavelength ends of these spectrometers often extend into the range where thermal emission dominates the measured flux (for the daytime surface temperatures of most asteroids). 
%For instance, the Rosetta and Dawn space missions carry two twin imaging spectrometers, VIRTIS \citep[][and references therein]{2011Sci...334..492C} and VIR \citep{DeSanctis2012Sci...336..697D}, working in the visible and in the near infrared spectral ranges between 0.5 and 5 \micron\ (it is important to bear in mind that these spectrometers are sensitive to temperatures $\gtrsim$ 170 K, which is the limiting temperature set by the noise equivalent spectral radiance).
%
%It is therefore desirable to derive the temperature of the surface of the asteroid from the thermal signal present in the wavelength range $\sim$ 3.5 - 5 \micron. Note that the \textbf{color temperature}, that is the temperature of an ideal black-body radiator produce a radiance of comparable hue to that of the asteroid surface, is derived from these measurements.  At these wavelengths, one cannot assume that the emissivity is close to 1.0, as is done at mid-infrared wavelengths.  Derivations of temperature from near-infrared spectral fluxes therefore have to separate temperature from spectral emissivity.  
%
At these wavelengths, one cannot assume that the emissivity is close to 1.0. It is therefore not practical to derive brightness temperatures.  Instead, the \textbf{color temperature} is derived, that is the temperature of a black body that emits with the same spectral shape.  Such derivations have to separate temperature from spectral emissivity.
The problem is under-constrained (N+1 unknowns, but only N data points), so there is no deterministic solution.  Spectral emissivities for fine-grained silicates trend in the same direction as the blackbody curve, so it would be very easy to mistake spectral emissivity variations for different temperatures.  The most statistically rigorous approach that has been applied to separating temperature and spectral emissivity in the 3 -- 5 \micron\ region is that of \cite{Keim2012Icar..221..395K} and \cite{Tosi2014Icar..240...36T} for Rosetta/VIRTIS data of Lutetia and Dawn/VIR data of Vesta.  
Temperatures thus measured
represent an average temperature in the field of view of a given pixel: illuminated hot zones and shadowed colder parts will both contribute.  
They do not directly correspond to a physical temperature of the soil; rather, they depend sensitively on the observation and illumination geometry
\citep[see][in particular their Fig.~9]{Rozitis2011MNRAS.415.2042R}, especially in the case of large illumination angles.

Microwave spectrometers such as MIRO \citep{Gulkis2007SSRv..128..561G} can provide both day and nightside thermal flux measurements. At sub-mm, mm, and longer wavelengths, asteroid soils become moderately transparent.  Subsurface layers contribute significantly to the observable thermal emission, thus providing information on the subsurface temperature.
Observable fluxes depend on the subsurface temperature profile, weighted by the wavelength-dependent electrical skin depths, so both a thermal and an electrical model are required to interpret such data \citep{Keim2012Icar..221..395K}.

We remind here that thermal infrared fluxes should be used as input data for TPMs and not (effective, color, or brightness) temperatures derived from radiometric methods, because of their dependence on illumination and observation angles !

\subsection{Sample-return missions}
\label{S:missions}
Space agencies across the planet are developing space missions to asteroids, notably sample-return missions to primitive (C and B type) 
near-Earth asteroids: Hayabusa-2, %\citep[launch scheduled in late 2014]{2011LPICo1611.5046Y} 
was launched by JAXA towards (162173) 1999 JU$_3$ on December 3, 2014,
and OSIRIS-REx is to be launched by NASA in 2016 \citep{Lauretta2012LPICo1667.6291L}. 
A good understanding of the expected thermal environment, which is governed by thermal inertia, is a key factor in planning spacecraft operations on or near asteroid surfaces.
E.g., OSIRIS-REx is constrained to sampling a regolith not hotter than \unit{350}{\kelvin}, severely constraining the choice of the 
latitude of the sample selection area on the body, the local time, and the arrival date on the asteroid. 

Both Hayabusa-2 and OSIRIS-REx are required to take regolith samples from the asteroid surface back to Earth. Obviously, this requires that regolith be present in the first place, which needs to be ascertained by means of ground-based thermal-inertia measurements. 
The sampling mechanism of OSIRIS-REx, in particular, requires relatively fine (\centi\metre-sized or smaller) regolith.   

\subsection{Accurate Yarkovsky and YORP modeling from TPMs}
\label{S:YarkoYORP}
%\cc{BEN !! }
Scattered and thermally emitted photons carry momentum.  Any asymmetry in the distribution of outgoing photons can, after averaging over an orbital period, impart 
a net recoil force (Yarkovsky effect) and/or a net torque (YORP effect) on the asteroid.  Both effects are more noticeable as the object gets smaller.
For small enough objects, the orbits can be significantly affected by the Yarkovsky effect, and their rotation state by YORP \citep{Bottke2006AREPS..34..157B,Vokrouhlicky2014astIV}.

The strength of the Yarkovsky effect is strongly influenced by thermal inertia \citep[][and references therein]{Bottke2006AREPS..34..157B} and by the degree of surface roughness \citep{Rozitis2012MNRAS.423..367R}. However, the strength and sign of the YORP rotational acceleration on an asteroid is independent of thermal inertia \citep{Capek2004Icar..172..526C}, but it is highly sensitive to the shadowing \citep{Breiter2009A&A...507.1073B}, self-heating \citep{Rozitis2013MNRAS.433..603R}, and surface roughness effects \citep{Rozitis2013MNRAS.433..603R} that are incorporated in thermophysical models. 

Accurate calculations of the instantaneous recoil forces and torques require an accurate calculation of surface temperatures as afforded by TPMs;
\citet{Rozitis2012MNRAS.423..367R,Rozitis2013MNRAS.433..603R} report on such models. %These instantaneous forces and torques are averaged over both the rotation and orbit to give the net effects acting on the asteroid, which can be compared against observations to gain additional information about the asteroid in question. 
Other than on thermal inertia, the Yarkovsky-induced orbital drift depends on the bulk mass density.  Therefore, Yarkovsky measurements combined with thermal-inertia measurements can be used to infer the elusive mass density
\citep{Mommert2014ApJ...786..148M,Mommert2014ApJ...789L..22M,Rozitis2014A&A...568A..43R,Rozitis2014Natur.512..174R,Rozitis2013A&A...555A..20R}. 
In the case of (101955) Bennu, the uncertainties in published values of thermal inertia \citep{Emery2014Icar..234...17E} and measured Yarkovsky drift  \citep{Chesley2014Icar..235....5C} are so small that the accuracy of the inferred mass density rivals that of the expected in-situ spacecraft result (1260 $\pm$ 70 kg m$^{-3}$, i.e., a nominal uncertainty of only ~6\%).
\citet{Rozitis2014Natur.512..174R} derived the bulk density of (29075) 1950 DA and used it to reveal the presence of cohesive forces stabilizing the object against the centrifugal force.

%A bulk density value can then be used to evaluate the asteroid's surface gravity environment, which can reveal the presence of cohesive forces if the asteroid is rotating fast enough  \citep[e.g., as seen on (29075) 1950 DA by][]{Rozitis2014Natur.512..174R}. 
In turn, the measured Yarkovsky drift can be used to infer constraints on  thermal inertia.  This was first done by \citet{Chesley2003Sci...302.1739C}  studying the Yarkovsky effect on (6489) Golevka and by \citet{Bottke2001Sci...294.1693B} studying the Koronis family in the main asteroid belt; 
both studies revealed thermal inertias consistent with expectations based on the observed correlation between thermal inertia and diameter (see \S~\ref{S:correlateGammaD}).

Whilst the YORP effect is highly sensitive to small-scale uncertainties in an asteroid's shape model \citep[][]{Statler2009Icar..202..502S}, it can be used to place constraints on the internal bulk density distribution of an asteroid \citep{Scheeres2008Icar..198..125S}. For instance, \cite{Lowry2014A&A...562A..48L} explain the YORP detection on (25143) Itokawa, which was opposite in sign to that predicted, by Itokawa's two lobes having substantially different bulk densities. However, unaccounted lateral heat conduction in thermal skin depth sized rocks could also explain, at least partially, this opposite sign result \citep{Golubov2012ApJ...752L..11G}.

%\subsection{Estimation of Asteroid physical properties by yarkovsky modeling}
%\cc{BEN + JOSH}
%Cite the work of Chesley on Golveka.\\
%Cite the work of Bootke on the Karins.\\
%Cite Emery et al. and Chesley et al for Bennu \cc{MAKE SHORT AS David V chapter is describing this}\\
%Cite Rozitis all papers 1950DA, Geographos, Icarus etc etc. 

\begin{small}
\begin{deluxetable}{r l c c c c c c c || r l c c c c c c c}
% !TEX root = astIVtpm_Rev2.tex
\tabletypesize{\scriptsize}
\tablecaption{\label{T:TI}Publised thermal inertia values}
\tablewidth{0pt}
\tablehead{Number & Name & $D$ & $\Delta_D$ & $\Gamma$ & $\Delta_\Gamma$ & Tax & $r$ & Ref. & Number & Name & $D$ & $\Delta_D$ & $\Gamma$ & $\Delta_\Gamma$ & Tax & $r$ & Ref.\\
 &  & (km) & (km) & (SI) & (SI) &  & (au) &  &  &  & (km) & (km) & (SI) & (SI) &  & (au) & }
\startdata
1 & Ceres & 923 & 20 & 10 & 10 & C & 2.767 & 1 & 1620 & Geographos & 5.04 & 0.07 & 340 & 120 & S & 1.1 & 12 \\
2 & Pallas & 544 & 43 & 10 & 10 & B & 2.772 & 1 & 1862 & Apollo & 1.55 & 0.07 & 140 & 100 & Q & 1.0 & 13 \\
3 & Juno & 234 & 11 & 5 & 5 & S & 2.671 & 1 & 2060 & Chiron & 142 & 10 & 4 & 4 & B/Cb & 8-15 & 14 \\
4 & Vesta & 525 & 1 & 20 & 15 & V & 2.3 & 2 & 2060 & Chiron & 218 & 20 & 5 & 5 & B/Cb & 13 & 15 \\
16 & Psyche & 244 & 25 & 125 & 40 & M & 2.7 & 3 & 2363 & Cebriones & 82 & 5 & 7 & 7 & D & 5.2 & 16 \\
21 & Lutetia & 96 & 1 & 5 & 5 & M & 2.8 & 4 & 2867 & Steins & 4.92 & 0.4 & 150 & 60 & E & 2.1 & 17 \\
22 & Kalliope & 167 & 17 & 125 & 125 & M & 2.3 & 5 & 2867 & Steins & 5.2 & 1 & 210 & 30 & E & 2.1 & 18 \\
32 & Pomona & 85 & 1 & 70 & 50 & S & 2.8 & 6 & 8405 & Asbolus & 66 & 4 & 5 & 5 & - & 7.9 & 19 \\
41 & Daphne & 202 & 7 & 25 & 25 & Ch & 2.1 & 7 & 3063 & Makhaon & 116 & 4 & 15 & 15 & D & 4.7 & 16 \\
44 & Nysa & 81 & 1 & 120 & 40 & E & 2.5 & 6 & 10199 & Chariklo & 236 & 12 & 1 & 1 & D & 13 & 14 \\
45 & Eugenia & 198 & 20 & 45 & 45 & C & 2.6 & 5 & 10199 & Chariklo & 248 & 18 & 16 & 14 & D & 13 & 13 \\
87 & Sylvia & 300 & 30 & 70 & 60 & P & 2.7 & 5 & 25143 & Itokawa & 0.32 & 0.03 & 700 & 100 & S & 1.1 & 8 \\
107 & Camilla & 245 & 25 & 25 & 10 & P & 3.2 & 5 & 25143 & Itokawa & 0.320 & 0.029 & 700 & 200 & S & 1.1 & 20 \\
110 & Lydia & 93.5 & 3.5 & 135 & 65 & M & 2.9 & 6 & 29075 & 1950~DA & 1.30 & 0.13 & 24 & 20 & M & 1.7 & 21 \\
115 & Thyra & 92 & 2 & 62 & 38 & S & 2.5 & 6 & 33342 & 1998~WT$_{24}$ & 0.35 & 0.04 & 200 & 100 & E & 1.0 & 8 \\
121 & Hermione & 220 & 22 & 30 & 25 & Ch & 2.9 & 5 & 50000 & Quaoar & 1082 & 67 & 6 & 4 & - & 43 & 15 \\
130 & Elektra & 197 & 20 & 30 & 30 & Ch & 2.9 & 5 & 54509 & YORP & 0.092 & 0.010 & 700 & 500 & S & 1.1 & 8 \\
277 & Elvira & 38 & 2 & 250 & 150 & S & 2.6 & 6 & 55565 & 2002~AW$_{197}$ & 700 & 50 & 10 & 10 & - & 47 & 22 \\
283 & Emma & 135 & 14 & 105 & 100 & P & 2.6 & 5 & 90377 & Sedna & 995 & 80 & 0.1 & 0.1 & - & 87 & 23 \\
306 & Unitas & 56 & 1 & 180 & 80 & S & 2.2 & 6 & 90482 & Orcus & 968 & 63 & 1 & 1 & - & 48 & 13 \\
382 & Dodona & 75 & 1 & 80 & 65 & M & 2.6 & 6 & 99942 & Apophis & 0.375 & 0.014 & 600 & 300 & Sq & 1.05 & 24 \\
433 & Eros & 17.8 & 1 & 150 & 50 & S & 1.6 & 8 & 101955 & Bennu & 0.495 & 0.015 & 650 & 300 & B & 1.1 & 25 \\
532 & Herculina & 203 & 14 & 10 & 10 & S & 2.772 & 1 & 101955 & Bennu & 0.49 & 0.02 & 310 & 70 & B & 1.1 & 26 \\
617 & Patroclus & 106 & 11 & 20 & 15 & P & 5.9 & 9 & 136108 & Haumea & 1240 & 70 & 0.3 & 0.2 & - & 51 & 27 \\
694 & Ekard & 109.5 & 1.5 & 120 & 20 & - & 1.8 & 6 & 162173 & 1999~JU$_3$ & 0.87 & 0.03 & 400 & 200 & C & 1.4 & 28 \\
720 & Bohlinia & 41 & 1 & 135 & 65 & S & 2.9 & 6 & 175706 & 1996~FG$_3$ & 1.71 & 0.07 & 120 & 50 & C & 1.4 & 29 \\
956 & Elisa & 10.4 & 0.8 & 90 & 60 & - & 1.8 & 10 & 208996 & 2003~AZ$_{84}$ & 480 & 20 & 1.2 & 0.6 & - & 45 & 27 \\
1173 & Anchises & 136 & 15 & 50 & 20 & P & 5.0 & 11 & 308635 & 2005~YU$_{55}$ & 0.306 & 0.006 & 575 & 225 & C & 1.0 & 30 \\
1580 & Betulia & 4.57 & 0.46 & 180 & 50 & C & 1.1 & 8 & 341843 & 2008~EV$_5$ & 0.370 & 0.006 & 450 & 60 & C & 1.0 & 31 \\
\enddata
\smallskip
 
References: 
[1] \cite{Muller1998A&A...338..340M},
[2] \cite{Leyrat2012A&A...539A.154L},
[3] \cite{Matter2013Icar..226..419M},
[4] \cite{ORourke2012P&SS...66..192O},
[5] \cite{Marchis2012Icar..221.1130M},
[6] \cite{Delbo2009P&SS...57..259D},
[7] \cite{Matter2011Icar..215...47M},
[8] \cite{Mueller2007PhD},
[9] \cite{Mueller2010Icar..205..505M},
[10] \cite{Lim2011Icar..213..510L},
[11] \cite{Horner2012MNRAS.423.2587H},
[12] \cite{Rozitis2014A&A...568A..43R},
[13] \cite{Rozitis2013A&A...555A..20R},
[14] \cite{Groussin2004A&A...413.1163G},
[15] \cite{Fornasier2013A&A...555A..15F},
[16] \cite{Fernandez2003AJ...126...1563},
[17] \cite{Lamy2008A&A...487.1187L},
[18] \cite{Leyrat2011A&A...531A.168L},
[19] \cite{Fernandez2002AJ...123..1050},
[20] \cite{Muller2014PASJ...66...52M},
[21] \cite{Rozitis2014Natur.512..174R},
[22] \cite{Cruikshank2005ApJ...624...53L},
[23] \cite{Pal2012A&A...541L...6P},
[24] \cite{Muller2014A&A...566A..22M},
[25] \cite{Muller2012A&A...548A..36M},
[26] \cite{Emery2014Icar..234...17E},
[27] \cite{Lellouch2013A&A...557A..60L},
[28] \cite{Muller2011A&A...525A.145M},
[29] \cite{Wolters2011MNRAS.418.1246W},
[30] \cite{Muller2013A&A...558A..97M},
[31] \cite{Ali-Lagoa2014A&A...561A..45A},
Note: for Ceres, Pallas, Juno, and Herculina $r$ is assumed equal to the semimajor asxis of the orbit.
\end{deluxetable}

\end{small}
% !TEX root = astIVtpm_Rev2.tex
\section{LATEST RESULTS FROM TPMs}
\label{S:results}

In the \emph{Asteroid III} era,  thermal properties were known for only a few asteroids, i.e., (1) Ceres, (2) Pallas, (3) Juno, (4) Vesta, (532) Herculina from \cite{2002A&A...381..324M}, and (433) Eros from \cite{Lebofsky1979Icar...40..297L}. Since then, the number of asteroids with known thermal properties has increased steadily. We count 59 minor bodies with known value of $\Gamma$ (see Tab.~\ref{T:TI}). Of these, 16 are near-Earth asteroids (NEAs), 27 main-belt asteroids (MBAs), 4 Jupiter Trojans, 5 Centaurs, and 7 trans-Neptunian objects (TNOs). 
%Some of these objects, with different sizes, albedos, spectroscopic classes and with regoliths of different mineralogical nature and grain sizes, were visited or will be visited by spacecrafts, providing ground-truth for the properties derived from the application of TPMs to remote sensing thermal infrared data. 

These classes of objects present very different physical properties such as sizes, regolith grain size, average value of the thermal inertia, and composition. Other important differences are their average surface temperature due to their very different heliocentric distances and orbital elements. The illumination and observation geometry are also diverse for different classes of objects. For instance, for TNOs and MBAs the phase angle of observation from Earth and Earth-like orbits is typically between a few and a few tens of degrees, respectively.  On the other hand, NEAs can be observed under a much wider range of phase angles than can approach hundred degrees and more \citet[see also][]{Muller2002M&PS...37.1919M}. A special care should be used in these cases to explicitly calculate the heat diffusion in craters instead of using the approximation of Eq.~\ref{E:LagerrosApprox}.

\subsection{Ground truth from space missions to asteroids}
%Until the last decade the lack of images of asteroidal regolith posed a limitation to confirm the reliability of predictions of regoliths nature based on thermal inertia values. However, recently, 
%Several space missions flew-by, orbited, and landed on asteroids of different sizes, albedos, spectroscopic classes and with regoliths of different mineralogical nature and grain sizes. Asteroids targets of said space missions have been extensively studied by means of telescopic observations in the thermal infrared, from the ground and from space, in order to obtain the physical properties of these bodies. The values of the physical parameters of these asteroids obtained by TPM analysis could have been compared with the corresponding values derived from spacecraft \emph{in situ} data. 

A number of asteroids have been, or will be, visited by spacecraft, providing ground-truth for the application of TPMs to remote-sensing thermal-infrared data.

\textbf{(21) Lutetia}:
based on ground-based data and a TPM, \citet{Mueller2006A&A...447.1153M} measured Lutetia's effective diameter and $p_V$ to within a few percent of the later Rosetta result 
\citep{Sierks2011Sci...334..487S}.  Their thermal-inertia constraint ($\Gamma<50$~\tiu) was refined by 
\citet{ORourke2012P&SS...66..192O} based on the Rosetta shape model and more than 70 thermal-infrared observations obtained from the ground,  \emph{Spitzer, Akari}, and \emph{Herschel}: $\Gamma$ = 5~\tiu with a high degree of surface roughness.
\citet{Keim2012Icar..221..395K} used MIRO aboard Rosetta to obtain a surface thermal inertia $\lesssim$~30~\tiu.
The low thermal inertia can be explained by a surface covered in fine regolith; \citet{Gundlach2013Icar..223..479G} infer a regolith grain size of about 200~\micron. The study of the morphology of craters by \citet{Vincent2012P&SS...66...79V} indicates abundant, thick (600 m), and very fine regolith, %(see Fig.~\ref{F:regolithImages}), 
confirming the TPM results.

\begin{figure}[h]
% \epsscale{1.5}
\plotone{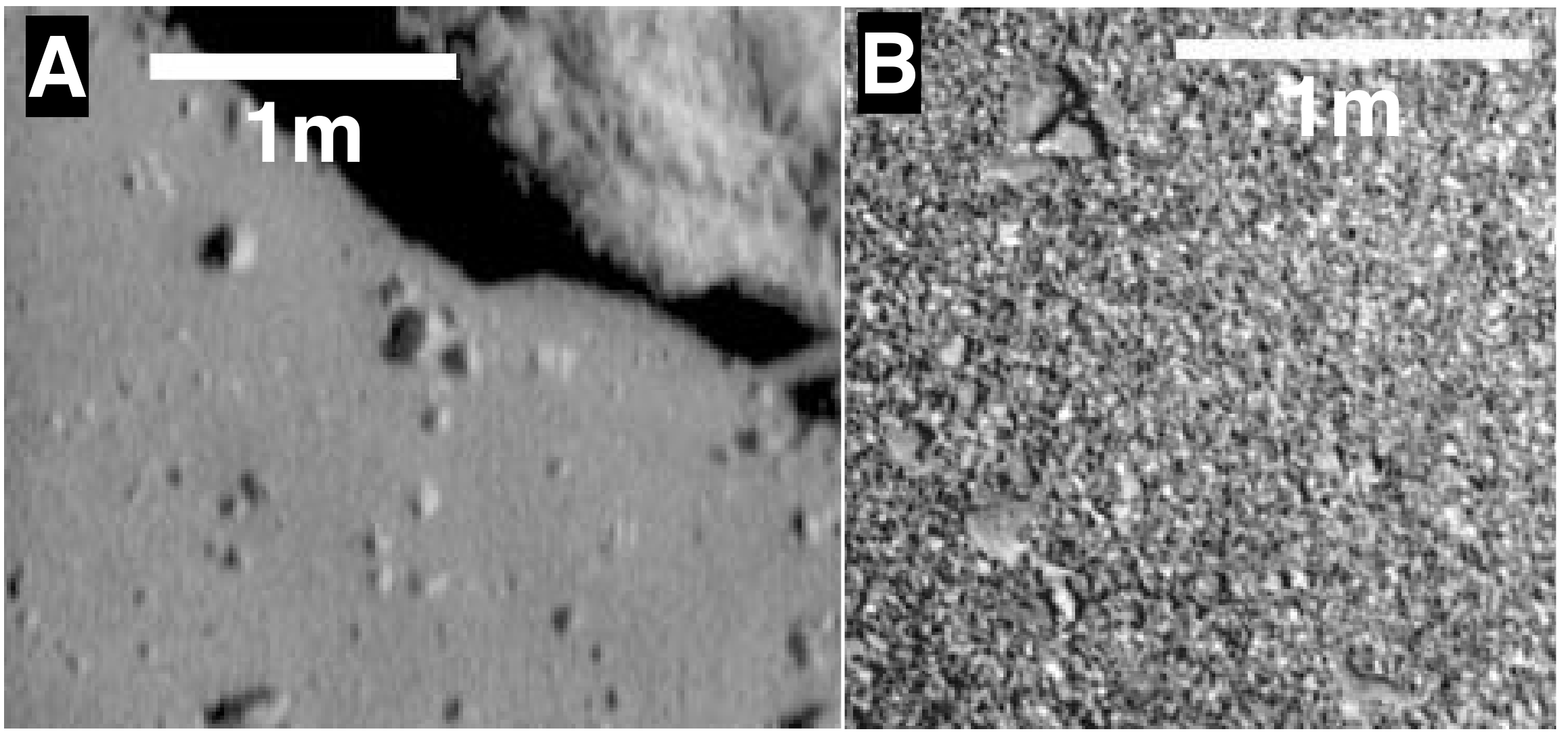}
\caption{\label{F:regolithImages}\small Higher $\Gamma$-values correspond to coarser regoliths. 
(A) Close-up image of (433) Eros from the NASA \emph{NEAR Shoemaker} mission reveals coarse regolith with grain size in the mm-range \citep[adapted from][]{Veverka2001Natur.413..390V}. The value of $\Gamma$ is $\sim$150 \tiu\ for Eros. 
%(B) Image of the soil of (21) Lutetia adapted from \citet{Sierks2011Sci...334..487S} from OSIRIS on board of the ESA mission Rosetta showing abundant, thick (600 m) and very fine regolith \citep{Vincent2012P&SS...66...79V}. The value of $\Gamma$ is $<$10 \tiu.
(B) Image from the JAXA \emph{Hayabusa} mission \citep[from][]{Yano2006Sci...312.1350Y} of the surface of (25143) Itokawa displaying gravel-like regolith.  The value of $\Gamma$ is $\sim$750 \tiu\ for Itokawa.
}  
\end{figure}

\textbf{(433) Eros } was studied by the NASA NEAR-Shoemaker space mission that allowed determination of the shape and size of this asteroid \citep[mean radius of 8.46 km with a mean error of 16 m;][]{Thomas2002Icar..155...18T}. \cite{Mueller2007PhD} performed a TPM analysis of the ground-based thermal infrared data by \cite{Harris1999Icar..142..464H}, obtaining a best-fit diameter of 17.8 km that is within 5\% of the \citet{Thomas2002Icar..155...18T} result of 16.9 km, and $\Gamma$  in the range 100 - 200 \tiu. The latter value, in agreement with TPM results of \citet{Lebofsky1979Icar...40..297L}, implies coarser surface regolith than that on the Moon and larger asteroids \citep[see, e.g.,][]{Mueller2007PhD,Delbo2007Icar..190..236D}. From the value of $\Gamma$ of \cite{Mueller2007PhD}, \cite{Gundlach2013Icar..223..479G} calculated a 1-3 mm typical regolith grain size for Eros. Optical images of the NEAR-Schoemaker landing site at a resolution of about 1 cm/pixel \citep{Veverka2001Natur.413..390V} show very smooth areas at the scale of the camera spatial resolution (Fig.~\ref{F:regolithImages}), likely implying mm or sub-mm grain size regolith, consistent with TPM results.

\textbf{(25143) Itokawa} physical properties were derived \emph{in-situ} by the JAXA sample-return mission Hayabusa, allowing us to compare the size, albedo and regolith nature derived from the TPMs with spacecraft results. \cite{Muller2014PASJ...66...52M} show an agreement within 2\% between the size and the geometric visible albedo inferred from TPM analysis of thermal-infrared data and the value of the corresponding parameters from Hayabusa data. The TPM thermal inertia value for Itokawa is around 750 \tiu, significantly higher than the value of our Moon (about 50 \tiu) and of other large main belt asteroids including (21) Lutetia, implying a coarser regolith on this small NEA. The corresponding average regolith grain size according to \cite{Gundlach2013Icar..223..479G} is $\sim\unit{2}{\centi\metre}$. 
Hayabusa observations from the optical navigation camera (ONC-T), obtained during the descent of the spacecraft to the ``Muses Sea'' region of the asteroid, reveal similar grain sizes, at a spatial resolution of up to 6 mm/pixel. In particular, \cite{Yano2006Sci...312.1350Y} describe ``Muses Sea'' as composed of numerous size-sorted granular materials ranging from several centimeters to subcentimeter scales.
Itokawa's regolith material can be classified as "gravel", larger than submillimeter regolith powders filling in ponds on (433) Eros (Fig.~\ref{F:regolithImages}). 

It is worth pointing out, however, that ``Muses Sea'' is not representative of Itokawa's surface as a whole.  Rather, it 
 was selected as a touchdown site because, in earlier Hayabusa imaging, it appeared as particularly smooth (minimizing operational danger for the spacecraft upon touch-down) and apparently regolith rich (maximizing the chance of sampling regolith). 
 Grain sizes measured at ``Muses Sea'' are therefore lower limits on typical grain sizes rather than values typical for the surface as a whole.

%While in the \emph{Asteroids~III} era there where only a handful of asteroids (and comets) for which we knew thermal inertia values, this number has grown considerably since then. 
%This is due to improvements in the TPMs, in the use of fast and parallel computers, and also in the availability of more accurate and novel thermal infrared observations.
%The \emph{Spitzer} and the \emph{WISE} telescopes, in particular, have opened a new era of asteroid thermal infrared observations and the exploitation of WISE data by thermophysical modeling has just begun \citep{Rozitis2014Natur.512..174R, Ali-Lagoa2014A&A...561A..45A}.

\begin{figure}[t]
% \epsscale{1.5}
%\plotone{figs/tiVSd.pdf}
%\includegraphics[trim=60 00 60 10,clip,scale=0.2]{DATA/Gamma_Diam_Original.pdf}
\includegraphics[trim=20 42 00 43,clip,scale=0.47]{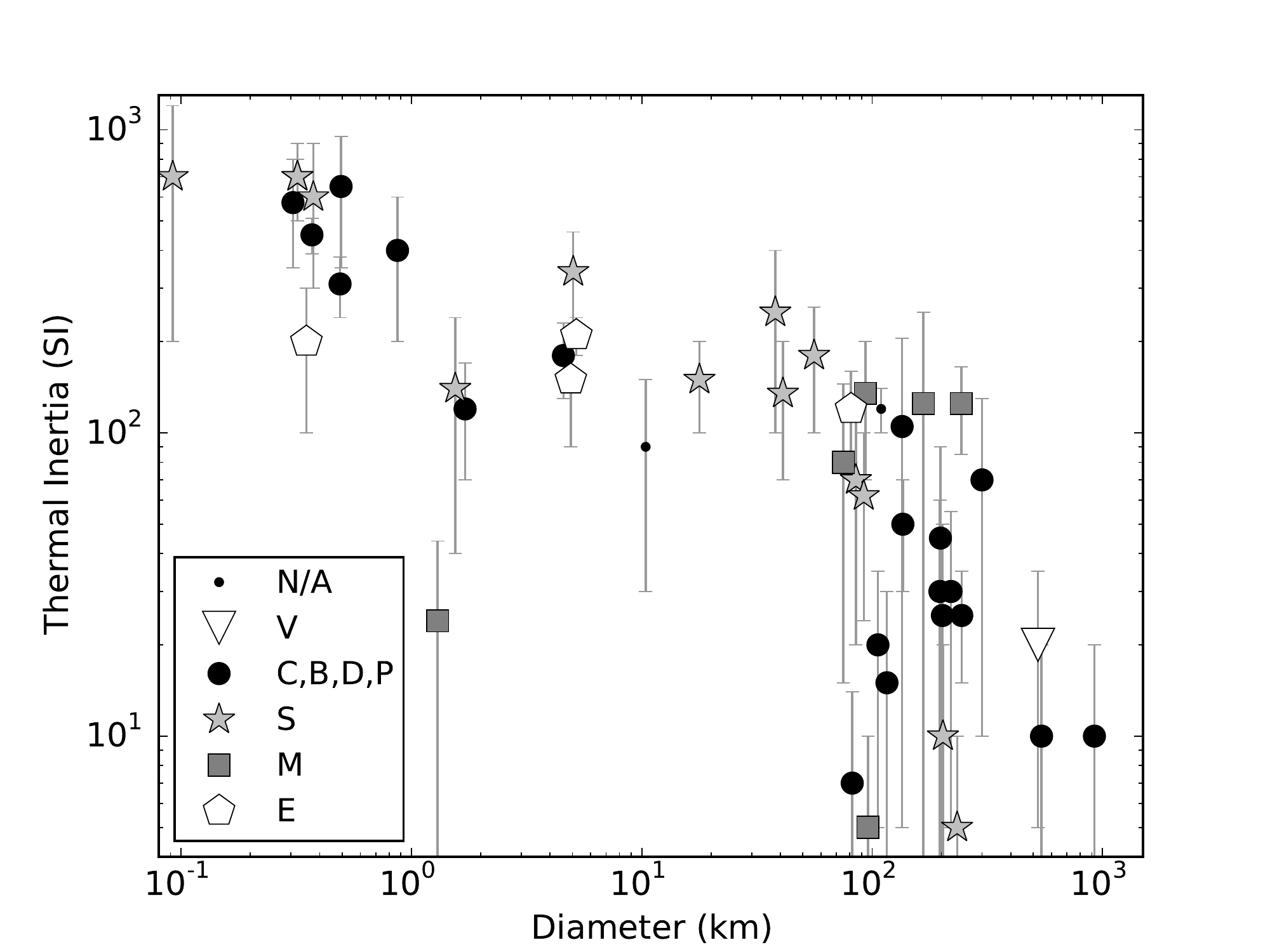}
\includegraphics[trim=20 00 00 43,clip,scale=0.47]{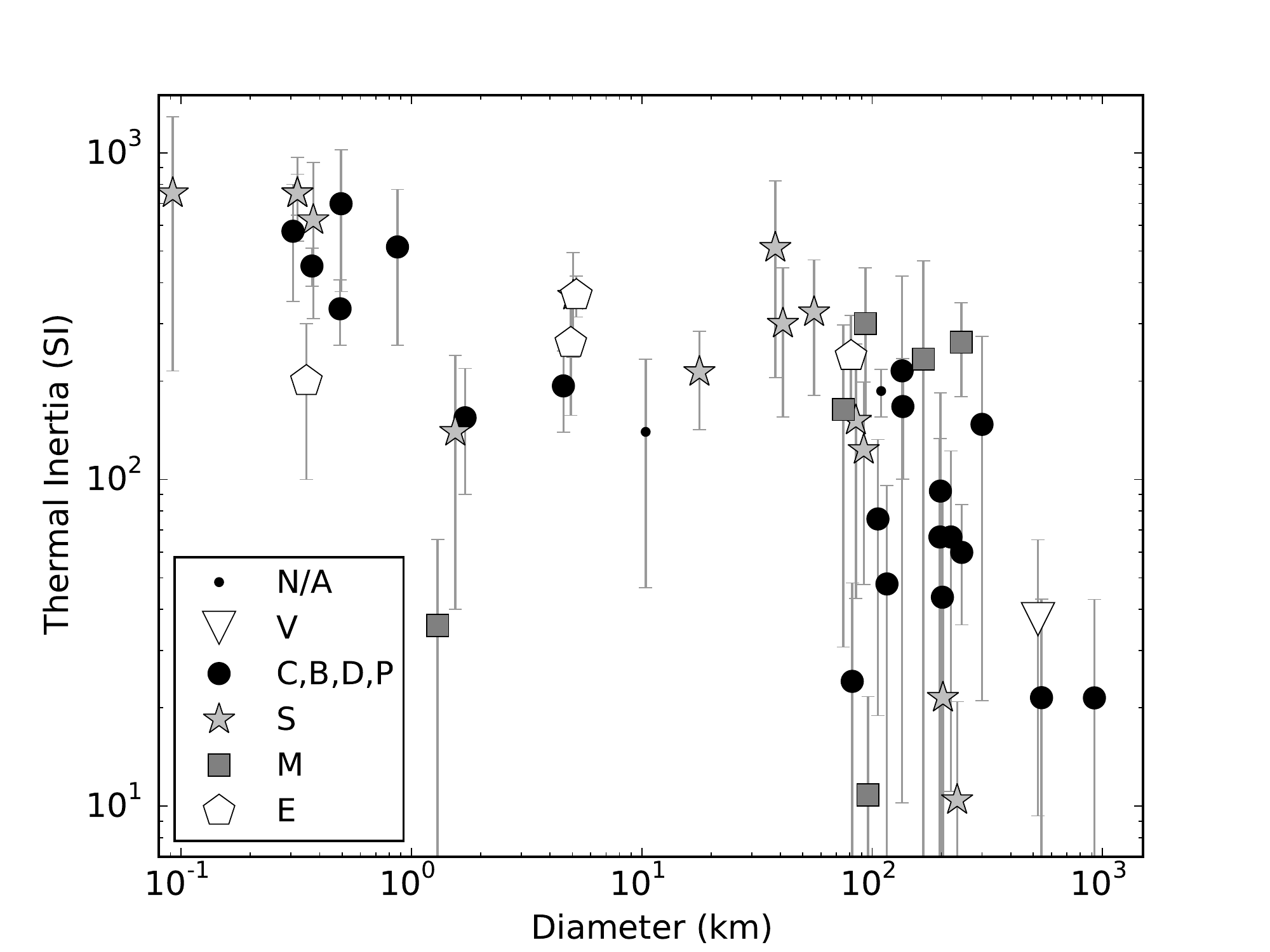}
\caption{\label{F:tiVSd} \small $\Gamma$ values vs. $D$ from Tab.~\ref{T:TI} for different taxonomic types (see key). 
%Trojans are also drawn with a thin-line circle around their main symbol. 
Top plot: original measurements, bottom plot: $\Gamma$ corrected to 1 au heliocentric distance for temperature dependent thermal inertia assuming Eq.~\ref{E:Gamma_r3/4} and the heliocentric distance at the time of thermal infrared observations reported in Tab.~\ref{T:TI}. Trojans, Centaurs and trans-Neptunian objects are not displayed.
% Missing Apophis from Muller
% Missing  Hanus et al. 2014 are to be added;
% Temperture correction TBD? 
% Change class of Ra-Shalom;
}  
\end{figure}

%\subsection{Low-$\Gamma$ for large and high-$\Gamma$ for small asteroids} 
%In the past, in the absence of images of asteroidal regoliths to be used as ground-truth for asteroid thermal inertia values, it was common practice to use the lunar $\Gamma_{\text{Moon}} \sim$ 50 \tiu\ as a reference. 

%However, novel data show that low-$\Gamma$-values are only valid for objects with D$\gtrsim$400 km, with asteroids in the size range between $\sim$ 300 and 50 km displaying thermal inertias in the range 1$\lesssim \Gamma \lesssim$ 100 \tiu.
%From the few early determination of asteroid thermal inertia values \citep{Muller1998A&A...338..340M}, it appeared clear that large asteroids (D$\gtrsim$100 km) have thermal inertia some factor lower than the lunar value ($\sim$50 \tiu), while smaller asteroids of few km in size have $\Gamma$-values several factors higher than lunar \citep{Harris2005Icar..179...95H,Mueller2007PhD,Delbo2007Icar..190..236D}, consistent with fine regolith on the former and more coarser grain-sized on the latter \citep{Gundlach2013Icar..223..479G},\cite[see also Tab.~\ref{T:TI}, Fig.~\ref{F:tiVSd}, and][and references therein]{Delbo2007Icar..190..236D}.

\subsection{Thermal inertia of large and small asteroids} 
\label{S:correlateGammaD}

%It is immediately obvious from Tab.~~\ref{T:TI} that $D>100$ km asteroids have low thermal inertia, of the order of the lunar thermal inertia ($\Gamma_{\text{Moon}} = 50$ \tiu) and below, while km-sized and smaller NEAs display much larger thermal inertia. 
%Following \cite{Gundlach2013Icar..223..479G}, the corresponding regolith grain sizes are in the mm-cm range for  km- and sub-km sized asteroids and between 10-100 \micron\ for $D>100$ km asteroids.

An inverse correlation between $\Gamma$ and $D$ was noticed by \cite{Delbo2007Icar..190..236D}, then updated \cite{Delbo2009P&SS...57..259D} and \cite{Capria2014GeoRL..41.1438C}.
This supported the intuitive view that large asteroids have, over many hundreds of millions of years, developed substantial insulating regolith layers, responsible for the low values of their surface thermal inertia. 
On the other hand, much smaller bodies, with shorter collisional lifetimes \citep[][and references therein]{Marchi2006AJ....131.1138M,Bottke2005Icar..175..111B},  have less regolith, and or larger regolith grains (less mature regolith), and therefore display a larger thermal inertia. 

In the light of the recently published values of $\Gamma$ (Tab.~\ref{T:TI}), said inversion correlation between $\Gamma$ and $D$ is less clear, in particular, when the values of the thermal inertia are temperature corrected (Fig.~\ref{F:tiVSd}). 
However, the $\Gamma$ vs $D$ distribution of $D>$ 100 km (large) asteroids is different than that of $D<$ 100 km (small) asteroids. Small asteroids typically have higher $\Gamma$-values than large asteroids, which  present a large scatter of $\Gamma$-values, ranging from a few to a few hundreds \tiu. This is a clear indication of a diverse regolith nature amongst these large bodies. A shortage of low $\Gamma$ values for small asteroids is also clear,  with the notable exception of 1950 DA, which has an anomalously low $\Gamma$-value compared to other NEAs of similar size \citep{Rozitis2014Natur.512..174R}.

Fig.~\ref{F:tiVSd} also shows previously unnoticed high-thermal-inertia C types, maybe related to CR carbonaceous chondrites, which contain abundant metal phases.  We also note that all E types in our sample appear to have a size-independent thermal inertia.

\subsection{Very low $\Gamma$-values}
We also note that the some of the C-complex outer main-belt asteroids and Jupiter Trojans have very low thermal inertia in the range between a few and a few tens of \tiu.
%Anyhow, the values of the thermal inertias of these small bodies are comparable with those of the four largest main belt asteroids \citep{Muller1998A&A...338..340M} and indicate that a very fine regolith covers these objects \cite{Gundlach2013Icar..223..479G}. 
In order to reduce the thermal inertia of a material by at least one order of magnitude (from the lowest measured thermal inertia of a meteorite, $\sim$650 \tiu\ at 200 K \citep{Opeil2010Icar..208..449O}, to the typical values for these large asteroids (Tab.~\ref{T:TI} and Fig.~\ref{F:tiVSd}), a very large porosity ($>$90\%) of the first few mm of the regolith is required \citep{Vernazza2012Icar..221.1162V}. This is consistent with the discovery that emission features in the mid-infrared domain (7--\unit{25}{\micron}, Fig.~\ref{F:SED_emissivity}) are rather universal among large asteroids and Jupiter Trojans \citep{Vernazza2012Icar..221.1162V}, and that said features can be reproduced in the laboratory by suspending meteorite and/or mineral powder (with grain sizes $<\unit{30}{\micron}$)
in IR-transparent KBr (potassium bromide) powder \citep{Vernazza2012Icar..221.1162V}. As KBr is not supposed to be present on the surfaces of these minor bodies, regolith grains must be "suspended" in void space 
%(see Fig.~\ref{F:vernazza12}) 
likely due to cohesive forces and/or dust levitation.
On the other hand, radar data  indicate a significant  porosity (40-50 \%) of the first $\sim$1 m of regolith \citep{Magri2001M&PS...36.1697M,Vernazza2012Icar..221.1162V}, indicating  decreasing porosity with increasing depth \citep[see Fig. 5 of][for a regolith schematics]{Vernazza2012Icar..221.1162V}.

%\begin{figure}[h]
%% \epsscale{1.5}
%%\includegraphics[trim=60 00 60 10,clip,scale=1.0]{figs/plotTIvsPHI.pdf}
%\plotone{figs/plotTIvsPHI.pdf}
%\caption{\label{F:tiVSphi} \small Porosity dependence of the value of the thermal inertia. In this example the case of Hawaiian basalts is considered. Data 
%adapted from \cite{Zimbelman1986Icar...68..366Z}. Line is the best fit of the model of \cite{Opeil2010Icar..208..449O} to the data of \cite{Zimbelman1986Icar...68..366Z}}  
%\end{figure}

%\begin{figure}[h]
%% \epsscale{1.5}
%\plotone{figs/vernazza12Fig5.pdf}
%\caption{\small  Schematic model of the asteroid surface structure as deduced from both the good correspondance between the KBr-diluted meteorite spectra and the asteroid spectra (for the first millimeter) and the low surface density inferred from radar measurements (for the first meter). The high porosity within the first millimeter may be due to cohesive forces (see grains on the upper right-hand side) and/or dust levitation (see grains on the upper left-hand side). From \cite{Vernazza2012Icar..221.1162V}.}  
%\label{F:vernazza12}
%\end{figure}

%{\color{red} TBD TBD The inferred surface properties can then be covered. }

\subsection{Average thermal inertia of asteroid populations}
\label{S:tiAverageFromEta}
As described before, the thermal inertia of an asteroid can be directly derived by comparing measurements of its thermal-infrared emission to model fluxes generated by means of a TPM. Typically, more than one observation epoch is required to derive the thermal inertia, in order to "see" the thermal emission from different parts of the asteroid's diurnal temperature distribution. Unfortunately, the large majority of minor bodies for which we have thermal-infrared observations have been observed at a single epoch and/or information about their gross shape and pole orientation is not available, precluding the use of TPMs.
However, if one assumes the thermal inertia to be roughly constant within a population of asteroids (e.g., NEAs) one can use observations of different asteroids under non-identical illumination and viewing geometries, as if they were from a unique object.  \cite{Delbo2003Icar..166..116D} noted that qualitative information about the average thermal properties of a sample of NEAs could be obtained from the distribution of the $\eta$-values of the sample as a function of the phase angle, $\alpha$. 
%In particular, the absence of large $\eta$-values (e.g., $\eta >$ 2) at small or moderate phase angles (e.g., $\alpha <$ 45$^\circ$), and the fact that $\eta$ tends to 0.8 for $\alpha$ approaching 0$^\circ$, was interpreted in terms of the NEAs having thermal inertia values similar of those of large main belt asteroids. 
\citet{Delbo2007Icar..190..236D} and \cite{Lellouch2013A&A...557A..60L} developed a rigorous statistical inversion method, based on the comparison of the distributions of published NEATM $\eta$-values vs $\alpha$, or vs. $r$ with that of a synthetic population of asteroids generated through a TPM, using realistic distributions of the input TPM parameters such as the rotation period, the aspect angle etc. \citet{Delbo2007Icar..190..236D} found that the average thermal inertia value for km-sized NEAs is around 200 \tiu. The average thermal inertia of binary NEAs is higher than that of non-binary NEAs, possibly indicating a regolith-depriving mechanism for the formation of these bodies \citep{Delbo2011Icar..212..138D}. The same authors also found that NEAs with slow rotational periods ($P>$10 h) have higher-than-average thermal inertia. 
From a sample of 85  Centaurs and trans-Neptunian objects observed with Spitzer/MIPS and Herschel/PACS, \cite{Lellouch2013A&A...557A..60L} found that surface roughness is significant, a mean thermal inertia $\Gamma= 2.5 \pm 0.5$ \tiu, and a trend toward decreasing $\Gamma$ with increasing heliocentric distance. The thermal inertias derived by \cite{Lellouch2013A&A...557A..60L} are 2-3 orders of magnitude lower than expected for compact ices, and generally lower than on Saturn's satellites or in the Pluto/Charon system. These results are suggestive of highly porous surfaces. 

\subsection{Relevant astronomical and laboratory data}
\label{S:tiLabData}

Physical interpretations of thermal-inertia estimates depend strongly on laboratory and ground-truth measurements of relevant material properties.  
%In practice, thermal models generally consider thermal inertia, $\Gamma$, rather than thermal conductivity, $\kappa$.  The reason is that, for relevant asteroidal materials and surface properties, heat capacity ($C$) and grain density ($\rho$) each span a fairly narrow range, whereas $\kappa$ can vary significantly. 
%
While in the \emph{Asteroid III} era, we based interpretation of thermal inertia on Earth analog materials, in the last few years laboratory measurements were performed on asteroid analog materials, i.e., meteorites. Meteorite grain densities range from $\sim$2800 \Du\ for CM carbonaceous chondrites to $\sim$3700 \Du\ for enstatite chondrites \citep{Consolmagno2006M&PS...41..331C,Macke2010M&PS...45.1513M,Macke2011M&PS...46..311M,Macke2011M&PS...46.1842M}. Heat capacities have been measured for a wide sampling of meteorites by \cite{Consolmagno2013P&SS...87..146C}, who find that values for stony meteorites are between 450 and 550 \Cu, whereas $C$ for irons tends to be smaller (330 -- 380 \Cu).  \cite{Opeil2012M&PS...47..319O,Opeil2010Icar..208..449O} present thermal conductivity measurements of stony meteorites, finding values of 0.5 \Ku\ for the carbonaceous chondrite Cold Bokkeveld to 5.5 \Ku\ for the enstatite chondrite Pillistfer.  Their one iron meteorite sample has a $\kappa$ of 22.4 \Ku.  They also find a linear correlation between  and the inverse of the porosity, from which \cite{Opeil2012M&PS...47..319O} conclude that the measured $\kappa$ of the samples is controlled more by micro-fractures than by composition.

\begin{figure}[h]
% \epsscale{1.5}
\plotone{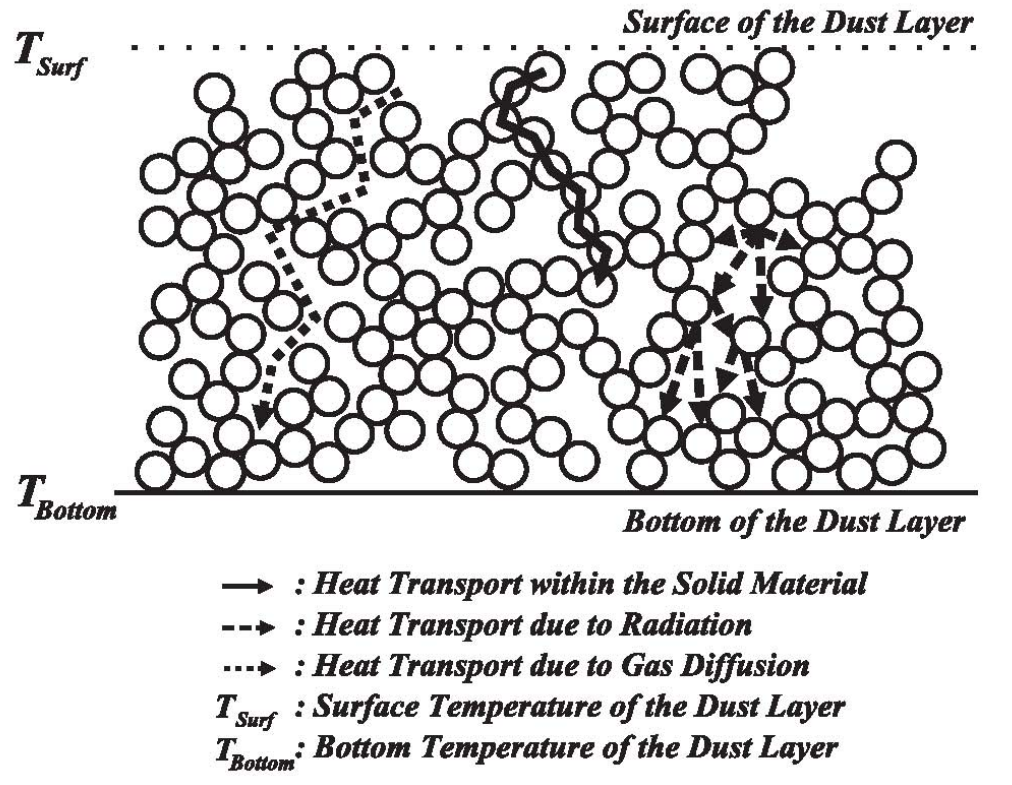}
\caption{\label{F:heatTransportModes} \small Diagram of the modes of heat transport in regoliths.  On airless bodies, heat can flow by conduction through grain boundaries (solid line) or by radiation between grains (dashed line).  The dotted line showing transport by gas diffusion is not relevant to asteroid surfaces. From \cite{Gundlach2012Icar..219..618G}.}  
\end{figure}

Grain size and packing, more than compositional heterogeneity, are responsible for different thermal inertias of different surfaces.  This also  explains why TPMs are capable of deriving asteroid physical parameters independently of the asteroid mineralogy. Conduction between grains is limited by the area of the grain contact \citep{Piqueux2009JGRE..114.9006P,Piqueux2009JGRE..114.9005P}.  As grain size decreases to diameters less than about a thermal skin depth (few cm on most asteroids), conduction is more and more limited \citep[e.g.,][]{Presley1997JGR...102.6551P}.  On bodies with atmospheres, conduction through the air in pores can often efficiently transport heat.  On airless bodies, however, radiation between grains, which is not very efficient, particularly at low T \citep[e.g.,][]{Gundlach2012Icar..219..618G}, is the only alternative to conduction across contacts (Fig.~\ref{F:heatTransportModes}).  Considering these two modes of energy transport and their dependence on grain size, \cite{Gundlach2013Icar..223..479G} developed an analytical approach for determining grain size from thermal inertia measurements.  They incorporated the measurements of material properties of meteorites measured above along with results of their own laboratory of heat transport in dusty layers.  
%This semi-empirical approach has large inherent uncertainties, but is easy to use and appears to be accurate.  
Additional laboratory measurements of conductivities of powdered meteorites under high vacuum would be valuable for more precise interpretation of asteroid thermal inertias.

%{\color{red} TBD TBD Include somewhere here this text -- LIKELY HERE -- }
%However, \citet{Gundlach2013Icar..223..479G} provided a calibration relation based on heat-transfer modeling in a granular medium.
%Basing themselves on a subset of the thermal-inertia data presented in Table\ \ref{T:TI},
%they find a typical regolith grain size in the \milli\metre--\centi\metre\ range for asteroids up to a diameter of $\sim\unit{100}{\kilo\metre}$, while larger objects show much finer regolith in the 10--\unit{100}{\micro\metre} range.

The classic opportunity for ground-truth thermal measurements came with the Apollo missions.  Astronauts on Apollo 15 and 17 carried out bore-hole style temperature measurements to depths of 1.4 m below the surface on Apollo 15 and 2.3 m below the surface on Apollo 17 \citep{Langseth1977,Vaniman199LunarSourceBook}. Thermal conductivity of about 0.001 \Ku\ was found in the top 2 to 3 cm of the lunar regolith, increasing to about 0.01 \Ku\ over the next few cm, then to values as high as 2 \Ku\ deeper into the surface where the regolith appears to have been very compacted \citep{Langseth1977}.  Low thermal inertias derived from remote thermal infrared measurements 
\citep[e.g.][]{Wesselink1948BAN....10..351W,Vasavada2012JGRE..117.0H18V} agree with the very low $\kappa$ in the topmost few cm of the lunar surface, and the Apollo measurements provide the necessary ground-truth for interpreting such low thermal inertias as very fine-grained, "fluffy" regolith. These measurements fostered, for instance, development of detailed models of lunar regolith \citep{Keihm1984Icar...60..568K}. Detailed thermal infrared observations and thermal models of the lunar regolith allows today estimating the subsurface rock abundance \citep[e.g.,][]{Bandfield2011JGRE..116.0H02B}, allowing geological studies of the regolith production rate.

\subsection{Dependence of $\Gamma$ with depth}
\label{S:tiDepth}
The depth dependence of typical asteroid regolith properties is poorly constrained at this point, which is why physical constants are typically assumed to be constant with depth.  MIRO observations of (21) Lutetia, however, showed the existence of a top layer with $\Gamma<30$ \tiu, while the thermal inertia of subsurface material appears to increase with depth much like on the Moon  \citep{Keim2012Icar..221..395K}. 

%Subsequent MIRO-based model predictions of the dayside surface temperatures reveal negative offsets of $\sim$5--30 K from the higher VIRTIS-measurements. By adding surface roughness in the form of 50\% fractional coverage of hemispherical mini-craters to the MIRO-based TPM, sufficient self-heating is produced to largely remove the offsets relative to the VIRTIS measurements and also reproduce the thermal limb brightening features (relative to a smooth surface model) seen by VIRTIS. Note that the limb brightening is predicted by \citet{Rozitis2011MNRAS.415.2042R}.

\subsection{Infrared limb brightening}
Recent modeling and observations show that, contrary to expectation, the flux enhancement measured in disk-integrated observations of the sunlit side of an asteroid \citep[e.g.,][]{Lebofsky1986Icar...68..239L} is dominated by limb surfaces rather than the subsolar region \citep{Rozitis2011MNRAS.415.2042R,Keim2012Icar..221..395K}. 
%This is clearly shown by the asteroid sunrise and sunset thermal infrared beaming enhancements being much greater than those at and near asteroid mid-day. 
This suggests that for the sunlit side of an asteroid, sunlit surfaces directly facing the observer in situations where they would not be if the surface was a smooth flat one are more important than mutual self-heating between interfacing facets raising their temperatures. Figure~9 of \citet{Rozitis2011MNRAS.415.2042R}  pictures this effect for a Gaussian random surface during sunrise viewed from different directions. The thermal flux observed is enhanced when viewing hot sunlit surfaces (i.e., Sun behind the observer), and is reduced when viewing cold shadowed surfaces (i.e., Sun in front of the observer).

\citet{Jakosky1990GeoRL..17..985J} also studied the directional thermal emission of Earth-based lava flows exhibiting macroscopic roughness. They found that enhancements in thermal emission were caused by viewing hot sunlit sides of rocks and reductions were caused by viewing cold shadowed sides of rocks. This agrees precisely with the model and adds further evidence that thermal infrared beaming is caused by macroscopic roughness rather than microscopic roughness.

The effect of limb brightening has also been measured from disk-resolved thermal infrared data ($<$5 \micron) acquired during sunrise on the nucleus of the comet 9P/Tempel 1 by the \emph{Deep Impact} NASA space mission \citep{Davidsson2013Icar..224..154D}, and from VIRTIS and MIRO measurements of the asteroid (21) Lutetia \citep{Keim2012Icar..221..395K}.

\subsection{Asteroid thermal inertia maps}
Disk-resolved thermal infrared observations, in the range between 4.5 -- 5.1 \micron, were provided by the instrument VIR \citep{DeSanctis2012Sci...336..697D} on board of the NASA DAWN \citep{Russell2012Sci...336..684R} spacecraft \citep[][and references therein]{Capria2014GeoRL..41.1438C}.  Form TPM analysis of VIR measurements, \citet{Capria2014GeoRL..41.1438C} obtained a map of the roughness and the thermal inertia of Vesta. 
The average thermal inertia of Vesta is 30 $\pm$ 10 \tiu, which is in good agreement with the values found by ground-based observations \citep{Muller1998A&A...338..340M,Chamberlain2007Icar..192..448C,Leyrat2012A&A...539A.154L}.
The best analog is probably the surface of the Moon, as depicted by \citet{Vasavada2012JGRE..117.0H18V} and \citet{Bandfield2011JGRE..116.0H02B}: a surface whose thermal response is determined by a widespread layer of dust and regolith with different grain sizes and density increasing toward the interior. Exposed rocks are probably scarce or even absent. \citet{Capria2014GeoRL..41.1438C} also show that Vesta cannot be considered uniform from the point of view of thermal properties. In particular, they found that the thermal inertia spatial distribution follows the global surface exposure age distribution, as determined by crater counting in \citet{Raymond2011SSRv..163..487R}, with higher thermal inertia displayed by younger terrains and lower thermal inertia in older soils. 

\citet{Capria2014GeoRL..41.1438C} also found higher-than-average thermal inertia terrain units located in low-albedo regions that contain highest abundance of OH, as determined by the 2.8 \micron\ band depth \citep{DeSanctis2012Sci...336..697D}. These terrains are associated with the dark material, thought to be delivered by carbonaceous chondrite like asteroids that have impacted Vesta at low velocity. Note that in general (carbonaceous chondrites) have lower densities and lower thermal conductivity \citep{Opeil2010Icar..208..449O} than basaltic material, which constitute the average Vestan terrain. This consideration would point to a lower thermal inertia rather than a higher one, as observed on Vesta. \citet{Capria2014GeoRL..41.1438C} conclude that the factor controlling the thermal inertia in these areas could be the degree of compaction of the uppermost surface layers, which is higher than in other parts of the surface.

\subsection{Thermal inertia of metal-rich regoliths}
In principle, the composition of the regolith and not only its average grain size and the degree of compaction also affects the thermal inertia of the soil \citep{Gundlach2013Icar..223..479G}. For instance iron meteorites have a higher thermal conductivity than ordinary and carbonaceous chondrites \citep{Opeil2010Icar..208..449O}. We thus expect that a metal iron rich regolith displays a higher thermal inertia than a soil poor of this component. \cite{Harris2014ApJ...785L...4H} compared values of the NEATM  $\eta$-parameter derived from WISE data with asteroid taxonomic classifications and radar data, and showed that the $\eta$-value appears to be a useful indicator of asteroids containing metal.
\citet{Matter2013Icar..226..419M} performed interferometric observations with  MIDI of the ESO-VLTI in thermal infrared of (16) Psyche
% This M-type asteroid has the second highest radar albedo amongst main belt asteroid\citet{Shepard2008Icar..195..184S}, which is indicative of a regolith with a high metal content. 
%Analysis of MIDI data 
and showed that Psyche has a low surface roughness and a thermal inertia value around 120 $\pm$ 40 \tiu, which is one of the higher values for an asteroid of the size of Psyche ($\sim$ 200 km). This higher than average thermal inertia supports the evidence of a metal-rich surface for this body.

% !TEX root = astIVtpm_Rev2.tex

\section{EFFECTS OF TEMPERATURES ON THE SURFACE OF ASTEROIDS}
\label{S:tempEffect}
%The sungrazing comet C/2012 S1 -- better known as Comet ISON --  spectacularly demonstrated the effect of extreme temperatures due to solar radiation on small bodies: as the comet was destroyed during its very close perihelion passage at 0.01244 AU on November 28, 2013. While that of Comet ISON is an extreme case with temperatures above the threshold for sublimation of silicates at that distance from the Sun, more moderate temperatures (and temperature variations) can effect important effects on asteroids. 

\subsection{Thermal cracking}
\label{S:thermalCracking}
The surface temperature of asteroids follows a diurnal cycle (see Fig.~\ref{F:diurnalTemperatureCurves}) with typically dramatic temperature changes as the Sun rises or sets.  The resulting, repeated thermal stress can produce cumulative damage on surface material due to opening and extension of microscopic cracks. This phenomenon is known as \emph{thermal fatigue} \citep{Delbo2014Natur.508..233D}.

Growing cracks can lead to rock break-up when the number of temperature cycles is large enough.  For typical asteroid properties, this process is a very effective mechanism for comminuting rocks and to form fresh regolith \citep{Delbo2014Natur.508..233D}. 
For \centi\metre-sized rocks on an asteroid 1 au from the Sun, thermal fragmentation is at least an order of magnitude faster than comminution by micrometeoroid impacts, the only regolith-production mechanism previously considered relevant \citep{Horz1997M&PS...32..179H,Horz1975Moon...13..235H}.

The efficiency of thermal fragmentation is dominated by the amplitude of the temperature cycles and by the temperature change rate \citep{Hall2001Geomo..41...23H}, which in turn depend on
heliocentric distance, rotation period, and the surface thermal inertia. The rate of thermal fragmentation increases with decreasing perihelion distance: at 0.14 au from the Sun, thermal fragmentation may erode asteroids such as (3200) Phaethon and produce the Geminids \citep{Jewitt2010AJ....140.1519J}, whereas in the outer Main Belt this process might be irrelevant. 
%For instance, the Geminids meteor stream parent body, the asteroid (3200) Phaethon, has a perihelion distance of only 0.14 AU. At that distance from the Sun, the temperature can reach about 1000 K and the typical diurnal temperature excursion of more than 400 K \citet{Delbo2014Natur.508..233D}.  \citet{Jewitt2010AJ....140.1519J} speculate that these extreme temperatures may cause break up of the surface material of this asteroid, providing allegedly a source of material for the annual Geminid meteor shower. 
%A potential observable implication of thermal weathering of rocks on asteroids come from the images of the surface of 
Thermal fragmentation of surface boulders is claimed by \citet{Dombard2010Icar..210..713D} to be source of fine regolith in the so-called "ponds" on the asteroid (433) Eros. Production of fresh regolith originating in thermal fatigue fragmentation may be an important process for the rejuvenation of the surfaces of near-Earth asteroids \citep{Delbo2014Natur.508..233D}. 
%and the prediction is that fresher asteroids are found at smaller heliocentric distances. Observations show that the fraction of NEAs with fresh surfaces, whose reflectance spectra resemble those of ordinary chondrites (Q-types), is increasing with decreasing perihelion distance with a dependence that mimics the curves of the amplitude of the temperature cycles of near-Earth asteroids \citep{Marchi2006MNRAS.368L..39M,Delbo2014Natur.508..233D}. 

Thermal cracking is reported on other bodies, too:
on Earth, particularly in super-arid environments \citep{Hall1999Geomo..31...47H,Hall2001Geomo..41...23H}, on the Moon \citep{Levi1973Metic...8..209L,Duennebier1974JGR....79.4351D}, Mercury \citep{Molaro2012JGRE..11710011M}, Mars \citep{Viles2010GeoRL..3718201V}, and on meteorites  \citep{Levi1973Metic...8..209L}.
Moreover, \citet{Tambovtseva1999P&SS...47..319T} suggest that thermal cracking could be an important process in the fragmentation and splitting of kilometer-sized comets
while in the inner solar system. 
Furthermore, \citet{Capek2010A&A...519A..75C} initially proposed that slowly rotating meteoroids or meteoroids that have spin vector pointing towards the sun can be broken up by thermal cracking. In a further development of their model, \citet{Capek2012A&A...539A..25C} showed that as the meteoroid approaches the Sun, the stresses first exceed the material strength at the surface and create a fractured layer. If inter-molecular forces \citep[e.g.,][]{Rozitis2014Natur.512..174R} are able to retain the surface layer, despite the competing effects of thermal lifting and centrifugal forces, the particulate surface layer is able to thermally shield the core, preventing any further damage by thermal stresses.

\subsection{Sun-driven heating of near-Earth asteroids and meteoroids}
It is known that heating processes can affect the physical properties of asteroids and their fragments, the meteorites \citep[see, e.g.,][]{Keil2000P&SS...48..887K}.

\begin{figure}[t]
% \epsscale{1.5}
\plotone{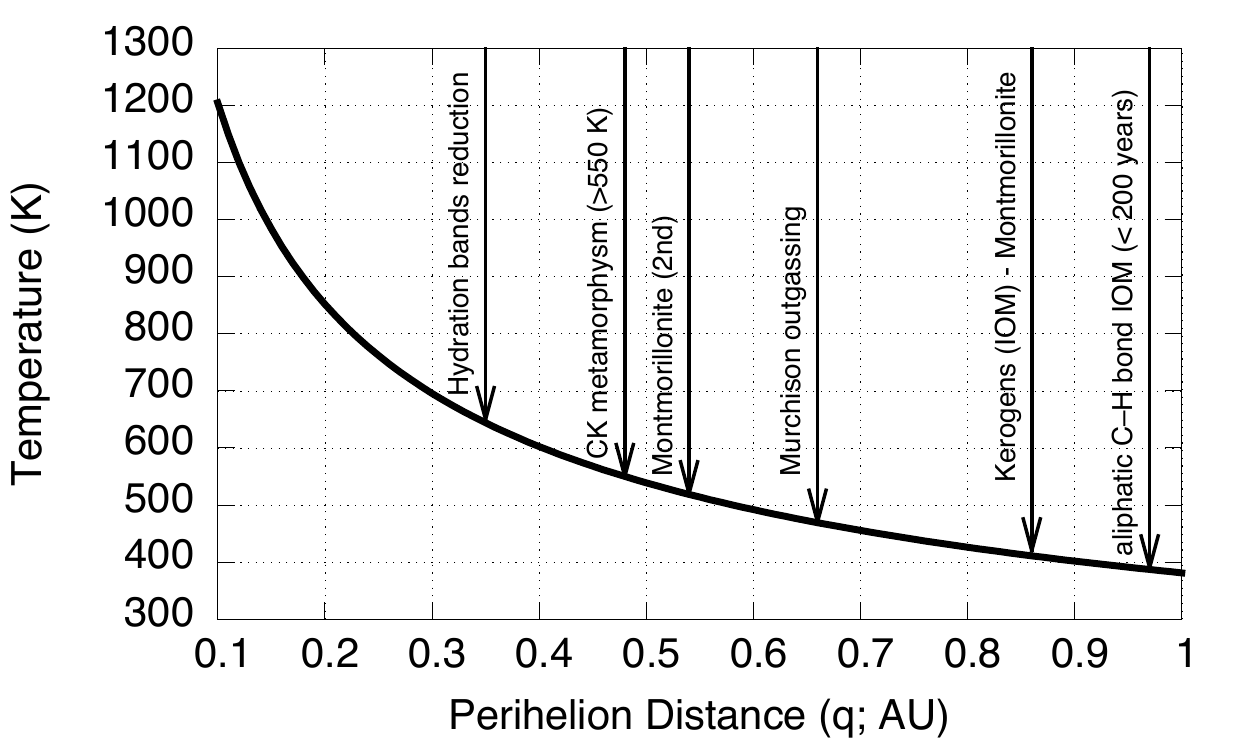}
\caption{\small Surface temperature of an asteroid or meteoroid as a function of the distance from the Sun. Vertical arrows indicate the threshold temperature for the thermal alteration/desiccation for a variety of chemical compounds discussed in the text \citep[see][and references therein for further information]{Delbo2011ApJ...728L..42D}. The temperature range for thermal metamorphism of the CK chondrites is from \cite{2012Icar..220...65C}.}  
\label{F:thermalEffects}
\end{figure}

Internal heating due to the decay of short-lived radionuclides was considered early on \citep{Grimm1993Sci...259..653G}.
\citet{Marchi2009MNRAS.400..147M} discuss close approaches to the Sun as an additional surface-altering heating mechanism.
In the present near-Earth asteroid population, the fraction of bodies with relatively small perihelion ($q$) is very small: about 1/2, 1/10, and 1/100 of the population of currently known near-Earth objects (11,000 as of the time of writing) have a perihelion distance below 1, 0.5, and 0.25 au, where maximum temperature are exceeding 400, 550, and 780 K, respectively (see Fig.~\ref{F:thermalEffects}). However, dynamical simulations show that a much larger fraction of asteroids had small perihelion distances for some time, hence experiencing episodes of strong heating in their past \citep{Marchi2009MNRAS.400..147M}. For instance, the asteroid  2004 LG was approaching the Sun to within only $\sim$5.6 solar radii some 3 ky ago, and its surface was baked at temperatures of 2500 K \citep{Vokrouhlicky2012A&A...541A.109V}.

Solar heating has a penetration depth  of typically a few \centi\metre\ \citep[see Eq.\ \ref{E:thermalSkinDepth} and][]{Spencer1989Icar...78..337S}. 
Organic components found on meteorites break up at temperatures as low as
300--670 K \citep[see Fig.~\ref{F:thermalEffects} and][]{Kebukawa2010M&PS...45...99K,Frost2000ThermoActa...346...63K,Huang1994AmeMineral...79...683K}, thus solar heating can remove these components from asteroid surfaces.

\subsection{Thermal metamorphism of meteorites}
Radiative heating from the Sun has been invoked as a mechanism for the thermal metamorphism of metamorphic CK carbonaceous chondrites \citep{2012Icar..220...65C}. The matrix of these chondrites shows textures consistent with a transient thermal event during which temperatures rose between 550  and 950 K. 
The inferred duration of these events is of the order of days to years, much longer than the time scale of shock events but shorter than the time scale for heating by the decay of radiogenic species such as $^{26}$Al \citep[e.g.,][]{Kallemeyn1991GeCoA..55..881K}.

\subsection{Subsurface ice sublimation}
Observational evidence for the presence of ice on asteroid surfaces stems from
the discovery of  main belt comets \citep[MBCs;][]{Hsieh2006Sci...312..561H}, the localized release of water vapor from the surface of (1) Ceres
% at a rate of about 6 kg/s
 \citep{Kuppers2014Natur.505..525K}, and the detection of spectroscopic signatures  interpreted as water ice frost on the surface of (24) Themis \citep{Rivkin2010Natur.464.1322R,Campins2010Natur.464.1320C} and of (65) Cybele \citep{Licandro2011A&A...525A..34L}.

The lifetime of ices on the surface and in the subsurface depends strongly on temperature.
TPMs have been used to estimate these temperatures.
This requires a modification of the ``classical'' TPM  as presented in section \ref{S:tpm}, such that heat conduction is coupled with  gas diffusion \citep{Schorghofer2008ApJ...682..697S,Capria2012A&A...537A..71C,Prialnik2009MNRAS.399L..79P}. The referenced models assume a spherical shape. 
As for the interior structure, 
\citet{Capria2012A&A...537A..71C,Prialnik2009MNRAS.399L..79P} assume a comet-like structure, i.e., an intimate mixture of ice and dust throughout the entire body, while \citet{Schorghofer2008ApJ...682..697S} consider  an ice layer underneath a rocky regolith cover. Sublimation of ice and the transport of water molecules through the fine-grained regolith is modeled in all cases. 

All authors agree that water ice exposed on asteroid surfaces sublimates completely on timescales much shorter than the age of the Solar System.  Therefore, asteroid surfaces were expected to be devoid of water ice, contrary to the observational evidence quoted above.
However, water ice can be stable over 4.5 Gy in the shallow subsurface, at a depth of $\sim$1--10 m. In particular, \citet{Fanale1989Icar...82...97F} showed that ice could have survived in the subsurface at the polar regions of Ceres. Large heliocentric distances, slow rotation, and a fine-grained regolith leading to low thermal conductivity and short molecular free path, all favor the  stabilization of subsurface water ice \citep{Schorghofer2008ApJ...682..697S}. The same authors conclude that rocky surfaces, in contrast to dusty surfaces, are rarely able to retain ice in the shallow subsurface.

To be observable on the surface, buried ice most be exposed.
\citet{Campins2010Natur.464.1320C} describe several plausible mechanisms such as impacts, recent change in the obliquity of the spin pole, and daily or orbital thermal pulses reaching a subsurface ice layer.

%\section{Temperature and spectral features}
%\cc{josh}
%High surface porosity as the origin of emissivity features in asteroid spectra [Vernazza  et al.] -- I have alstry describe in the Results from TPM section \\ 
%
%Taken from \cite{Lucey2002Icar..155..181L}\\
%In a series of papers, Roush and Singer (Roush 1984, Singer and Roush 1985, Roush and Singer 1986) and later Schade and Wasch (1999), Hinrichs (2000), and Moroz et al. (2000) showed that the near-infrared reflectance spectra of olivine and pyrox- ene powders vary with temperature. As these minerals are important constituents of many meteorites and asteroids, Roush and Singer (1987) and Moroz et al. (2000) discussed the implications of this phenomenon for spatially resolved spectral measurements of asteroids and showed that the effect should be detectable using reasonable assumptions regarding surface thermal properties. 

% !TEX root = astIVtpm_Rev2.tex

\section{FUTURE CHALLENGES FOR TPMs}
\label{S:future}

The \emph{Spitzer} and \emph{WISE} telescopes have opened a new era of asteroid thermal-infrared observations and the exploitation of their data through TPMs has just begun \citep[e.g.,][]{Ali-Lagoa2014A&A...561A..45A,Rozitis2014Natur.512..174R,Emery2014Icar..234...17E}. At the moment, the limiting factor is the availability of accurate asteroid shape models. However, optical-wavelength all-sky surveys such as PanSTARRS, LSST, and Gaia are expected to produce enormous photometric data sets leading to thousands of asteroid models. We envision the availability of thousands of thermal-inertia values in some years from now, enabling more statistically robust studies of thermal inertia as a function of asteroid size, spectral class, albedo, rotation period, etc. 

For instance, the distribution of $\Gamma$ within asteroid families will be crucial in the search of evidence of 
asteroid differentiation: asteroid formation models and meteorite studies suggest that hundreds of planetesimals experienced complete or partial differentiation. An asteroid family formed from the catastrophic disruption of such a differentiated asteroid should contain members corresponding to the crust, the mantel and the iron core. However, the observed spectra and albedos are very homogeneous across asteroid families.
Thermal inertia might help in separating iron rich from iron poor family members, supposedly originating respectively from the core and mantle of the differentiated parent body \citep[e.g.,][]{Matter2013Icar..226..419M,Harris2014ApJ...785L...4H}.

At any size range, Fig.~\ref{F:tiVSd} shows an almost  tenfold variability in thermal inertia, corresponding to difference in average regolith grain size of 
almost two orders of magnitude \citet{Gundlach2013Icar..223..479G}. 
%What could be the explanation for such different regoliths? 
%
For small near-Earth asteroids, this could be due to a combination of thermal cracking \citep{Delbo2014Natur.508..233D}, regolith motion \citep{Murdoch2014astIV}, and cohesive forces \citep{Rozitis2014Natur.512..174R}. Faster rotation periods allow more thermal cycles, which then enhances thermal fracturing. It also encourages regolith to move towards the equator where the gravitational potential is at its lowest \citep{Walsh2008Natur.454..188W}. And for the extremely fast rotators, large boulders/rocks could be selectively lofted away, because they stick less well to the surface than smaller particles.
For $D>100$ km sized asteroids, $\Gamma$-values might be help to distinguish between primordial and more recently re-accumulated asteroids. The former had $\sim$ 4 Gy of regolith evolution, the latter have a less developed and therefore coarser regolith.

The high-precision thermal-infrared data of WISE and Spitzer pose new challenges to TPMs, as model uncertainties are now comparable to the uncertainty of the measured flux. This will become even more important with the launch of the James Webb Space Telescope (JWST). In particular, the accuracy of the shape models might represent a limiting factor \citep[e.g.,][]{Rozitis2014A&A...568A..43R}. The next challenge will be to allow the TPM to optimize the asteroid shape. This seems to be possible as the infrared photometry is also sensitive to shape, provided good-quality thermal data are available \citep{Durech2014astIV}. 

New interferometric facilities, such as MATISSE, LBTI, and ALMA,  will become available in the next years requiring TPMs to calculate precise disk-resolved thermal fluxes \citep{Durech2014astIV}. The wavelengths of ALMA, similar to those of MIRO, will allow to measure the thermal-infrared radiation from the subsoil of asteroids, thus providing further information about how thermal inertia varies with depth.

%To date, the surface roughness derived from thermal infrared observations has been compared against that measured in-situ for the Moon only, e.g., by using the work of \citet{Helfenstein1999Icar..141..107H}. However, to make an accurate comparison, \citet{Rozitis2011MNRAS.415.2042R} suggest that the roughness needs to measured over a baseline as large as the shape model facets used (or the detector footprint in spatially resolved measurements) at thermal skin depth scales. For the Moon, this requires measuring roughness at $\sim$1 cm scales over a $\sim$10 km baseline, which can be approximated by combining RMS slopes measured at different spatial scales within this range in quadrature. 

Certainly, constraining roughness is one of the future challenges for TPMs. To do so from disk-integrated data requires a range of wavelengths and solar phase angles. Low phase angle measurements are enhanced by beaming whilst high phase angle measurements are reduced by beaming. In particular, shorter wavelengths are affected more than longer wavelengths. 

Moreover, the future availability of precise sizes and cross sections of asteroids from stellar occultation timing \citep{Tanga2007A&A...474.1015T}, combined with shape information derived from lightcurve inversion \citep{Durech2014astIV} will allow to remove the need to constrain the object size from TPM analysis. Iinfrared fluxes will thus be converted into highly reliable thermal inertia and roughness values.

%$\bullet$ Improved computational speed (GPUs from graphic cards) 
%$\bullet$ Large data sets: WISE, NEOCam using asteroid poles and shapes from Gaia, LSST, PANSTARSS, etc.
%$\bullet$ Interferometry (ALMA?), High spatial resolution (ELTs ?).
%$\bullet$ Global inversion (see chapter by Durech et al.).
%$\bullet$ Support of near surface spacecraft operations: e.g., find fine regolith. 

\section*{Acknowledgments} We are grateful to S. Green and to an anonymous referee for their thorough reviews. MD thanks J. Hanus and the support from the French Agence National de la Recherche (ANR) SHOCKS.

\bibliographystyle{my_apalike_A4}
%\bibliographystyle{elsarticle-harv}
%\bibliography{biblio,bibliography_all,mypapers}
\bibliography{mypapers}

\end{document}